\PassOptionsToPackage{unicode}{hyperref}
\PassOptionsToPackage{hyphens}{url}
\documentclass[
]{article}
\usepackage{setspace}
\usepackage{iftex}
\ifPDFTeX
  \usepackage[T1]{fontenc}
  \usepackage[utf8]{inputenc}
  \usepackage{textcomp} 
\else 
  \usepackage{unicode-math} 
  \defaultfontfeatures{Scale=MatchLowercase}
  \defaultfontfeatures[\rmfamily]{Ligatures=TeX,Scale=1}
\fi
\usepackage{lmodern}
\ifPDFTeX\else
\fi
\IfFileExists{upquote.sty}{\usepackage{upquote}}{}
\IfFileExists{microtype.sty}{
  \usepackage[]{microtype}
  \UseMicrotypeSet[protrusion]{basicmath} 
}{}
\usepackage{xcolor}
\usepackage[margin=1in]{geometry}
\usepackage{graphicx}
\makeatletter
\def\maxwidth{\ifdim\Gin@nat@width>\linewidth\linewidth\else\Gin@nat@width\fi}
\def\maxheight{\ifdim\Gin@nat@height>\textheight\textheight\else\Gin@nat@height\fi}
\makeatother
\setkeys{Gin}{width=\maxwidth,height=\maxheight,keepaspectratio}
\makeatletter
\def\fps@figure{htbp}
\makeatother
\makeatletter
 \let\@cite@ofmt\@firstofone
 \def\@biblabel#1{}
 \def\@cite#1#2{{#1\if@tempswa , #2\fi}}
\makeatother
\newlength{\cslhangindent}
\newlength{\csllabelwidth}
 {\begin{list}{}{%
  \setlength{\itemindent}{0pt}
  \setlength{\leftmargin}{0pt}
  \setlength{\parsep}{0pt}
  \ifodd #1
   \setlength{\leftmargin}{\cslhangindent}
   \setlength{\itemindent}{-1\cslhangindent}
  \fi
  \setlength{\itemsep}{#2\baselineskip}}}
 {\end{list}}
\usepackage{calc}

\usepackage{float}

\usepackage[backend=bibtex, style=authoryear, maxnames=2, maxbibnames=5, sorting=nyt, giveninits=false, dashed=false, url=false, doi=false, eprint=false]{biblatex}
\usepackage{amsmath}
\usepackage{amsthm}
\usepackage{amssymb}
\usepackage{hyperref}
\usepackage{booktabs}
\usepackage{longtable}
\usepackage{soul}
\renewcommand{\arraystretch}{0.9}
\usepackage{array}
\usepackage{enumitem}
\usepackage{arydshln} 
\usepackage{multicol} 
\usepackage{caption}
\usepackage{pdflscape} 
\usepackage{lscape}
\usepackage{adjustbox}
\usepackage{placeins}
\usepackage{bbm}
\usepackage[hang,flushmargin]{footmisc}
\usepackage{multirow}
\usepackage{wrapfig}
\usepackage{colortbl}
\usepackage{threeparttable}
\usepackage{threeparttablex}
\usepackage[normalem]{ulem}
\usepackage{makecell}
\usepackage{caption}
\ifLuaTeX
  \usepackage{selnolig}  
\fi
\usepackage{bookmark}
\IfFileExists{xurl.sty}{\usepackage{xurl}}{} 
\hypersetup{
  pdftitle={Multivariate Rough Volatility},
  pdfauthor={Ranieri Dugo; Giacomo Giorgio; Paolo Pigato},
  hidelinks,
  pdfcreator={LaTeX via pandoc}}

\title{Multivariate Rough Volatility}
\author{Ranieri Dugo\footnote{Department of Economics and Finance,
  University of Rome Tor Vergata,
  \href{mailto:dugo.ranieri@students.uniroma2.eu}{\nolinkurl{ranieri.dugo@students.uniroma2.eu}}} \and Giacomo
Giorgio\footnote{Department of Mathematics, University of Rome Tor
  Vergata,
     \href{mailto:giorgio@mat.uniroma2.it}{\nolinkurl{giorgio@mat.uniroma2.it}}
  } \and Paolo Pigato\footnote{Department of Economics and
  Finance, University of Rome Tor Vergata,
    \href{mailto:paolo.pigato@uniroma2.it}{\nolinkurl{paolo.pigato@uniroma2.it}}}} 
\date{\today}

\begin{document}
\maketitle
\setstretch{1}
\renewcommand{\thefootnote}{} 
\footnotetext{{\bf Acknowledgements:} We are grateful to Tommaso Proietti, Stefano De Marco, Alessandro Casini, Davide Pirino, Mathieu Rosenbaum, Mikkel Bennedsen, to the participants to the 4th Meeting in Probability in Rome, the QFFE24 in Marseille, the VI Aarhus Workshop in Econometrics, the 3rd Edition of the Finance and Business Analytics Conference for discussion. 
We thank J.-F. Coeurjolly for sharing the code for simulating the multivariate fractional Brownian Motion. 
We also thank the two anonymous referees for their valuable guidance in improving the manuscript.
Funding: PP was supported by the project PRICE, financed by the Italian Ministry of University and Research under the program PRIN 2022, Prot. 2022C799SX. RD and PP were supported by the project E83C25000470005 financed by University of Rome Tor Vergata.
}
\renewcommand{\thefootnote}{\arabic{footnote}}

\begin{abstract}
\noindent
Motivated by empirical evidence from the joint behavior of realized volatility time series, we propose to model the joint dynamics of log-volatilities using a multivariate fractional Ornstein-Uhlenbeck process. This model is a multivariate version of the Rough Fractional Stochastic Volatility model introduced in Gatheral, Jaisson, and Rosenbaum, \emph{Quant. Finance}, 2018. It allows for different Hurst exponents in the different marginal components and non trivial interdependencies.

We discuss the main features of the model and propose a Generalized Method of Moments estimator that jointly identifies its parameters. We derive the asymptotic theory of the estimator and perform a simulation study that confirms the asymptotic theory in finite sample.

We conduct an extensive empirical investigation of all realized-volatility time series covering the entire span of about two decades in the Oxford-Man realized library, and of a small spot-volatility system. Our analysis shows that these time series are strongly correlated and can exhibit asymmetries in their empirical cross-covariance function, accurately captured by our model. These asymmetries lead to spillover effects, which we derive analytically within our model and compute based on empirical estimates of model parameters. Moreover, in accordance with the existing literature, we observe behaviors close to non-stationarity and rough trajectories. 
\end{abstract}


\newtheorem*{theorem}{Theorem}
\newtheorem{prop}{Proposition}
\newtheorem{remark}{Remark}
\newtheorem*{assumptions}{Assumption}
\newpage

\section{Introduction}\label{introduction}

We consider the log-normal fractional model proposed in \textcite{CR98} and \textcite{GJR18} for volatility time series and introduce a multivariate extension in which the dynamics of the log-volatility are specified by a multivariate fractional Ornstein-Uhlenbeck process. This is a system of stochastic equations coupled through the noise terms, which are given by a multivariate fractional Brownian motion (mfBm) in the sense of \textcite{ACLP10,lav09}. We introduced and analyzed this process in \textcite{dgp1}.  Its marginal components are one dimensional fractional Ornstein-Uhlenbeck (fOU) processes, consistently with the established literature on fractional and rough volatility, with Hurst regularity parameter that can be different across components. The cross-covariance between each two components is ruled by two additional parameters, invisible to the univariate dynamics, one related to the contemporaneous correlation, and another related to the time reversibility of the process. Both are inherited from the driving mfBm.

The fact that empirical log-volatility is mean-reverting, Gaussian, and exhibits fractional features is well established (\cite{fouque2000mean, andersen2001distribution, ding1996modeling}). A mean-reverting, fractional, Gaussian volatility model, able to reproduce the long-memory behavior of volatility, and based on the fOU process (see \cite{cheridito2003fractional}), was first introduced in \textcite{CR98}. Estimations of this model on empirical data, on short time lags, have subsequently hinted to small Hurst parameters for volatility time series, supporting the ``rough volatility'' specification by \textcite{GJR18}. The empirical rough behavior of the trajectories of this model has been confirmed by a number of econometrical studies (\cite{bolko2023gmm,WXY23,ES20,BIANCHI2023}), and the estimation of similar models also points to rough behavior of the volatility time series (\cite{BLP21,CHLRS2,CHLRS1,FTW,LMPR}). The rough volatility specification, moreover, is supported by the option pricing literature with particular attention to the implied volatility skew (\cite{bayer2016pricing, BFGHS,FGP22,LMPR,del.dem.seg}, see also \cite{GuyonSkew}). Concerning the modelling of multivariate time series and their interdependence structure, let us mention \textcite{bibinger2025modeling,PDHGS06, PDHGS01, WPHS11, PS}, where applications to economics, but also to physics, physiology, and genomics, are considered. 
Time-reversibility has also been widely considered in the financial literature, at least in the unidimensional setting (see e.g. \cite{zumbach2009time, cordi}). 

\smallskip
\noindent
\textbf{Contribution}: We consider a multivariate system of volatility time series and assume that their joint dynamics follows a multivariate fractional Ornstein-Uhlenbeck (mfOU) process. To the best of our knowledge, this is the first continuous-time fractional volatility model in a multivariate setting.\newline\indent
We estimate the parameters of the model using a Generalized Method of Moments procedure (\cite{hansen1982large}), which aims at matching model-implied and empirical cross-covariances.\newline\indent
First, we derive the asymptotic theory of the estimator, proving speed of convergence and asymptotic normality, and perform an extensive simulation study that confirms these results in finite sample. We also propose a variation tailored to the ``slow mean reversion'' case. This method builds on previous results from \textcite{dgp1}, where the mfOU process is defined and its cross-correlations parameters are estimated with a method of moments, assuming known univariate parameters. Here we propose and analyze a more efficient estimation procedure, able to estimate simultaneously all the parameters of the multivariate system, with the goal of modeling a multivariate realized volatilty time series.\  
Then, we estimate the model on 20 years of 22 realized volatility time series data from the Oxford-Man library, finding a remarkably good fit, particularly with respect to cross-covariances and their possibly asymmetric decay as a function of the time lag. In the model, this decay is determined by the mean-reversion and time-reversibility parameters, which determine the asymmetry degree, and by the sum of the Hurst exponents. The presence of asymmetries in the empirical cross covariances suggests the presence of spillover effects between realized volatility time series, which we derive analytically employing within our model ideas from \textcite{diebold2012better}. Then, we evaluate these quantities with the parameter estimates obtained in sample. This spillover analysis might capture volatility dynamics better and in a simpler fashion than previous models, as it is shown in \textcite{bibinger2025modeling} that traditional models may not perform well when applied to forecasting fractional time series. We also show that our model can be used to capture intraday volatility dynamics by fitting a small spot-volatility system for US indices.  Our empirical analysis provides additional evidence in favour of the rough volatility specification and of behaviors close to non-stationarity of volatility time series.

In the spirit of the present paper, \textcite{WXY23} considers the fractional OU process for log-realized volatility, focusing on the univariate case and the forecasting problem. GMM estimation in the univariate case is used in \textcite{bolko2023gmm}, modeling spot volatility and approaching the inference problem starting from moments of integrated variance. Here, we mainly use realized volatility as our measure of volatility because it allows us to model global markets and, therefore, do not focus in depth on estimating the roughness of the spot-volatility process, which has already been extensively studied in the recent literature. 
In a recent, very related paper, \textcite{bibinger2025modeling} propose modeling and forecasting multivariate realized volatility with the mfBM, in the time reversible case, which is supported empirically in their dataset.

\smallskip
\noindent
{\bf Outline:} Section \ref{the-model} introduces and describes the model. Section \ref{estimation-method} discusses parameter identification and volatility measurement. Section \ref{monte-carlo-study} evaluates with simulations the finite-sample performance of the estimator. Section \ref{empirical-analysis} presents the empirical estimation results. Section \ref{spillovers} examines spillover effects. Section \ref{conclusion} concludes. Proofs and supplementary results are provided in the Appendix, with additional empirical analyses and working code available online.

\section{The Model}\label{the-model}

The model we propose for the logarithm of volatility is a multivariate version of the Rough Fractional Stochastic Volatility model by \textcite{GJR18}. It is the solution to a mean-reverting stochastic differential equation (SDE) driven by the mfBm by \textcite{ACLP10}, which we call multivariate fractional Ornstein-Uhlenbeck (mfOU) process,. We introduced this continuous-time process in \textcite{dgp1} and we briefly outline it below, together with the mfBm.  

The mfBm, $(W_t^H)_t=(W_t^{H_1},\dots,W_t^{H_N})_t,$ is a vector-valued Gaussian process in dimension \(N\) governed by the parameters \(H\in(0,1)^N\), \(\rho\in[-1,1]^{N\times N}\), and \(\eta\in\mathbb{R}^{N\times N}\). In this process, \(H=(H_1,\dots,H_N)\) is the vector of Hurst exponents, which are allowed to be different across components and determine the memory of the process and the roughness of its trajectories. These properties are directly transferred to the mfOU process. The matrix \(\rho\) represents the contemporaneous correlation of the mfBm at each point in time. One generic entry is $\rho_{i,j}=\mathbb{E}[W_t^{H_i}W_t^{H_j}]$ and it satisfies $\rho_{i,j}=\rho_{j,i}$ and $\rho_{i,i}=1$. The matrix \(\eta\) is related to the properties of interdependence in time and time reversibility of the process. The matrix $\eta$ is antisymmetric, i.e. $\eta_{i,j}=-\eta_{j,i}$, and $\eta_{i,i}=0$. The parameter $\eta$ determines the asymmetry in the cross-covariance function, indeed $\mathbb{E}[W_t^{H_i}W_s^{H_j}]\ne\mathbb{E}[W_t^{H_j}W_s^{H_i}]$ if and only if $\eta_{i,j}\ne0$. In addition, the parameters $H,\ \rho, \ \text{and}\ \eta$ need to satisfy admissibility constraints in order to have a well-defined covariance function (see \cite{dgp1}). An introduction to the mfBm for econometricians is given in \textcite{bibinger2025modeling}.
In our model, we add component-wise mean-reverting dynamics to the mfBm as follows. The mfOU process, \(\left(Y_t\right)_t=(Y_t^1,\dots,Y_t^N)_t\), is the vector-valued process whose components solve pathwise (\cite{cheridito2003fractional}) the equations 
\begin{equation}
dY_t^i=\alpha_i(\mu_i-Y_t^i)dt + \nu_idW_t^{H_i},\ \ \ \ \ i = 1, \dots, N,
\label{eq:langevin}
\end{equation} where \(\mu_i\in\mathbb{R}\) is the long-term mean,
\(\nu_i>0\) is the diffusion coefficient, and \(\alpha_i>0\) is the
speed of mean reversion. Let us write $\nu=\left(\nu_1,\dots,\nu_N\right)$ and $\alpha=\left(\alpha_1, \dots,\alpha_N\right)$. Equation (\ref{eq:langevin}) has a stationary solution given by 
\begin{equation}
Y_t^i = \mu_i+\nu_i\int_{-\infty}^t e^{-\alpha_i\left(t-s\right)}dW_s^{H_i}, \ \ \ \ \ i = 1, \dots, N.
\label{def:mfou}
\end{equation} 
This is an autoregressive process, as one can see from the following representation of the solution to \eqref{eq:langevin}.
\begin{equation*}
Y_t^i = e^{-\alpha_i\Delta} Y_{t-\Delta}^i+\left(1-e^{-\alpha_i \Delta}\right)\mu_i+\nu_i\int_{t-\Delta}^t e^{-\alpha_i\left(t-s\right)}dW_s^{H_i}, \ \ \ \ \ i = 1, \dots, N.
\end{equation*} 
In this work, we assume to observe the log-realized volatility, described by the mfOU process, at the stationary regime. Its expectation is
\begin{equation*}
\mathbb{E}\left[Y_t\right]=\mu=(\mu_1,\dots,\mu_N).
\end{equation*} 
Let us denote
\(\gamma_{i,j}(k):= \text{Cov}(Y_{t+k}^i,Y_{t}^j),\ k\in\mathbb{R}\), the
cross-covariance function. We assume here \(H_{i,j}=H_i+H_j\ne 1\), in which case 
\begin{equation}
\gamma_{i,j}(k)=e^{-\alpha_i k}\gamma_{i,j}(0)+\nu_i\nu_j e^{-\alpha_i k}H_{i,j}(H_{i,j}-1)\frac{\rho_{i,j}+\eta_{i,j}}{2}\int_0^ke^{\alpha_i v}\left(\int_{-\infty}^0e^{\alpha_j u}(v-u)^{H_{i,j}-2}du\right)dv,
\label{eq:ccf}
\end{equation} 
where the covariance, \(\gamma_{i,j}(0)\), is 
\begin{equation*}
\gamma_{i,j}(0)=\frac{\Gamma(H_{i,j}+1)\nu_i\nu_j}{2(\alpha_i+\alpha_j)}\left(\left(\alpha_i^{1-H_{i,j}}+\alpha_j^{1-H_{i,j}}\right)\rho_{i,j}+\left(\alpha_j^{1-H_{i,j}}-\alpha_i^{1-H_{i,j}}\right)\eta_{i,j}\right).
\end{equation*} 
In \textcite{dgp1} we showed that this cross-covariance decays as a power law with exponent \(H_{i,j}-2\), as \(k\to\infty\), therefore allowing for long-range interdependence (meaning that the cross-covariance is not integrable) when \(H_{i,j}>1\), in analogy to long-memory in the univariate case. 
Note that a different expression holds for the covariance function when \(H_{i,j}=1\), which corresponds to a discontinuity of the cross-covariance as a function of the Hurst exponents.  However, this is of little relevance for the modeling of volatility, since typically $H_i\ll1/2$, which implies $H_{i,j}\ll1$, and moreover the case $H_{i,j}=1$ has zero-measure in the space of parameters\footnote{It corresponds to a manifold with a lower dimension than the full space of parameters}. We also stress that while roughness and long-term memory properties are inherited from the underlying mfBm, the local behavior of the process is influenced by the mean-reversion introduced by the drift. In addition, the asymmetry in the cross-covariance function of the mfOU process is jointly determined by $\eta$ and$\alpha$. Indeed, to have $\gamma_{i,j}(k)=\gamma_{j,i}(k)$ we need both $\eta_{i,j}=0$ and $\alpha_i=\alpha_j$.
 
In line with the univariate case treated by \textcite{GJR18}, if the mean reversion coefficients are small, the process behaves locally as a mfBm.

\begin{prop}
Let $\left(W_t^H\right)_t$ be a mfBm, $\left(Y_t\right)_t$ be the mfOU process in \eqref{def:mfou}. Then, as $\alpha\to 0$,
\begin{equation*}
\mathbb{E}\left[\sup_{t\in[0,T]}\left|\left|Y_t-Y_0-\nu\odot W_t^H\right|\right|\right]\to0,
\end{equation*}
where $|| \cdot ||$ represent the $L^2$ norm and $\odot$ indicates the Hadamard product.
\end{prop}

\emph{Proof.} Follows from the univariate result in \textcite{GJR18}.

Note that $\gamma_{i,j}(k)$ and $\gamma_{i,j}(0)$ depend on $\alpha_i$ and $\alpha_j$. In the regime of slow mean reversion, the cross-covariance function of the mfOU process is approximately linear in $k^{H_{i,j}}$.

\begin{prop} For $H_{i,j}<1$ and fixed $k>0$, as $(\alpha_i, \alpha_j)\to(0,0)$,
\begin{equation*}
\gamma_{i,j}(k) = \gamma_{i,j}(0) -\frac{\rho_{i,j}+\eta_{i,j}}{2}\nu_i\nu_jk^{H_{i,j}} + o(1).
\label{eq:ccfl}
\end{equation*}
\end{prop}

\textit{Proof}. Follows by taking the limit in
(\ref{eq:ccf}) and standard computations.

An analogous approximation result holds taking $k\to 0$ instead of $(\alpha_i, \alpha_j)\to(0,0)$ (see \cite[Lemma 2.5]{dgp1}).

\section{Estimation method}\label{estimation-method}

\subsection{Parameter identification}

In order to jointly estimate all the parameters in our model, we rely on the Generalised Method of Moments (GMM, \cite{hansen1982large}). The number of parameters that need to be estimated for a system of dimension $N$ amounts to \(p=N(N+2)\), which includes \(N \times 3\) parameters governing the marginal distributions, specifically \(\alpha_i\), \(\nu_i\), and \(H_i\) for \(i = 1, \dots, N\), and \(N(N - 1)/2\) parameters in each of the matrices \(\rho\) and \(\eta\), determining the multivariate dynamics. We subtract the sample mean from the observations of the process in the beginning, so we do not deal with \(\mu_i\).

We define the parameter vector \(\theta = (\alpha_i, \nu_i, H_i, \rho_{i,j}, \eta_{i,j}, \ i = 1, \dots, N,\ i<j<N) \in \Theta \subset \mathbb{R}_+^N \times \mathbb{R}_+^N \times (0,1)^N \times [-1,1]^{N(N-1)/2} \times \mathbb{R}^{N(N-1)/2}\), which represents the full set of parameters to be estimated.
While maximizing the likelihood function would be the most efficient approach, the non-Markovian nature of the process makes this computationally infeasible. In \textcite{amblard2011identification}, discrete filtering techniques are used to estimate the mfBm. Here, we adopt a 2-step GMM procedure.  

In the GMM approach, we consider an overdetermined system of equations in the parameters to be estimated, which we call moment conditions. The estimator is defined as the parameter value that is closest, in a mean-square sense, to solving the system. Let \((Y_{i\Delta})_{i=0}^n\) be a set of \(n\in\mathbb{N}\) discrete observations over the interval \([0,T]\), observed at time intervals \(\Delta=T/n\). Our goal is for the model-implied cross-covariances, given in Equation \eqref{eq:ccf}, and calculated with the estimated parameters, \(\gamma_{i,j}^k(\hat\theta_n)\) (see definition below), to be as close as possible to sample cross-covariances \[ \hat\gamma_{i,j;n}^k=\frac{1}{n-k}\sum_{\ell=1}^{n-k}\left(Y_{\ell+k}^i-\hat\mu_i\right)\left(Y_{\ell}^j-\hat\mu_j\right), \] where \(\hat\mu_i=\frac{1}{n}\sum_{\ell=1}^n Y_\ell^i,\ i=1,\dots,N\) are  computed in our framework before the optimization takes place.

We define the vectors that contain, respectively, model and sample cross-covariances, ordered consistently, in an obvious manner:
\begin{align*}
    \gamma(\theta) &= \left(\left(\gamma_{ij}^k(\theta)\right)_{k\in\mathcal{L},\ i,j=1,\dots,N}\right)^T 
    \in \mathbb{R}^{N(L+(N-1)(L-1/2))}, \\
    \hat\gamma_n &= \left(\left(\hat\gamma_{i,j;n}^k\right)_{k\in\mathcal{L},\ i,j=1,\dots,N}\right)^T 
    \in \mathbb{R}^{N(L+(N-1)(L-1/2))},
\end{align*} 
where \(L\) is the cardinality of the set of indices \(\mathcal{L}\subset\mathbb{N}\cup\{0\}\). This set can be appropriately chosen, depending on the problem under consideration. See \textcite{andersen1996gmm} for a general discussion, or Section \ref{finite-sample-performance} and \textcite{bolko2023gmm} for our specific case.

We define the GMM estimator \(\hat\theta_n\) as the value of \(\theta\) that minimizes the loss function
\begin{equation*}
\mathcal{T}_n\left(\theta\right)=
\left(\hat\gamma_n-\gamma(\theta)\right)^T \mathcal{W}_n \left(\hat\gamma_n-\gamma(\theta)\right),
\end{equation*}
i.e.
\begin{equation}
\hat\theta_n=\arg\min_{\theta}\mathcal{T}_n\left(\theta\right),
\label{eq:min}
\end{equation} 
where \(\mathcal{W}_n\) is a symmetric, positive semidefinite matrix of order \(N(L+(N-1)(L-1/2))\), possibly data dependent, that converges to a constant as $n\to\infty$. Stationarity and ergodicity of the mfOU process, together with the regularity of the cross-covariance function, imply the following.
\begin{prop}  
Let $\theta_0\in\Theta$ be the true value of the parameter in a population distributed as the mfOU process. If $H_{i,j}\ne 1,\ \forall i,j=1,\dots,N$, $\mathcal{W}_n\to \mathcal{W},\ as\ n\to\infty$, with initial condition for the parameter optimization in a suitable neighborhood of $\theta_0$, then
\begin{enumerate}[label=\Roman*.]
\item the GMM estimator is consistent, i.e. 
$$
\hat\theta_n \overset{p}{\to}\theta_0\ \ \ \ \ \ \text{as}\ n\to\infty,
$$
\item when $\max_{1\le i \le N}{H_i}<\frac{3}{4}$, the GMM estimator is asymptotically normal, i.e.
$$
\sqrt{n}\left(\hat\theta_n-\theta_0\right)\overset{d}{\to}N(0,\Sigma)\ \ \ \ \ \ \text{as}\ n\to\infty,
$$
where $\Sigma=\left(J_\gamma^T\mathcal{W}J_\gamma\right)^{-1}J_\gamma^T\mathcal{W}\Gamma W J_\gamma\left(J_\gamma^T\mathcal{W}J_\gamma\right)^{-1}$, $J_\gamma:=J_\gamma(\theta_0)
=-\partial\gamma(\theta)/\partial\theta|_{\theta_0}$ is the Jacobian matrix of $\hat\gamma_n-\gamma\left(\theta\right)$ at $\theta_0$, and $\Gamma$ is the asymptotic covariance matrix of $\hat\gamma_n$.  
\end{enumerate}
\label{theo:mde}
\end{prop}

The proof of the above results builds on the usual theory for GMM estimators and the convergence of sample cross-covariances to population ones (cf. Appendix A). We expect, but do not discuss here, a non-Gaussian result to hold if \(\max_{1\le i \le N}H_i> {3}/{4}\), similarly to Theorem 3.9 in \textcite{dgp1}.

\begin{remark}
The assumption $H_{i,j}\ne 1$ is an identification assumption. Global identification requires $\mathbb{E}\left[\hat\gamma_n-\gamma(\theta)\right]=0$ if and only if $\theta=\theta_0$, which is hard to verify. Instead, the concept of local identification, upon which we rely and which coincides with global identification if the moment conditions are linear in the parameters, requires the simpler conditions of (i) continuous differentiability of $\hat\gamma_n-\gamma(\theta)$ around $\theta=\theta_0$, and (ii) $\mathbb{E}\left[\nabla_\theta\left(\hat\gamma_n-\gamma(\theta_0)\right) \right]$ or a consistent estimate of it, having full rank. In our optimization problem, local identification ensures a well-behaved search space in a neighborhood of the solution. Local identifiability can indeed be verified aside from the discontinuity point of the cross-covariance function, $H_{i,j}\ne 1$, by seeing that the Jacobian matrix, $J_\gamma$, has full column rank.
\end{remark}

\begin{remark} 
The inequality
\begin{equation*}
(J_\gamma^T\mathcal{W}J_\gamma)^{-1}J_\gamma^T\mathcal{W}\Gamma \mathcal{W}D(J_\gamma^T\mathcal{W}J_\gamma)^{-1} - (J_\gamma^T\Gamma^{-1}J_\gamma)^{-1}\ge0,
\end{equation*}
suggests that the highest efficiency for $\hat\theta_n$ could be achieved by choosing $\mathcal{W}=\Gamma^{-1}$, which delivers the lowest asymptotic variance, $\Sigma^*=(J_\gamma^T\Gamma^{-1}J_\gamma)^{-1}$. However, the matrix $\Gamma$ is not explicit within our model and can be difficult to estimate if we deal with data which are persistent or close to persistent, as is often the case with realized volatility, or high-dimensional. In addition, the optimal GMM estimator is known to deliver lower standard errors at the cost of a higher bias in the parameter of mean reversion (\cite{andersen1996gmm}). Therefore, we proceed with a two step GMM procedure that performs the first step with the identity matrix, \(\mathcal{W}_n:=I\), and the second step with the inverse of the diagonal part of the Newey-West (\cite{nw87}) estimate of $\Gamma$, $\mathcal{W}_n=\text{diag}(\hat\Gamma)^{-1}$. Additional iterations do not seem to improve the estimation.  
\end{remark}

We solve the optimization in (\ref{eq:min}) numerically, using initial conditions provided by a combination of the univariate estimators from \textcite{WXY23} with those from \textcite{dgp1} for the correlation parameters, which take the former as given. This initial condition can be reasonably expected to be located near the true parameter value, $\theta_0$, as required by the local identification assumption, but is however inefficient and does not consider the overall estimation error. We perform the numerical optimization with the L-BFGS-B algorithm within the \emph{optim()} function in R 4.3.3.

\subsection{Volatility measurement}\label{volatility-measurement}

In the rough volatility literature researchers have divided their attention between spot and integrated volatility measures. In an asset pricing model where prices $(S_t)_{t>0}$ follow the stochastic differential equation 
$$
dS_t=S_t\sigma_t dW_t,
$$ 
it is customary to refer to $\sigma_t$ as spot volatility, to
$$
IV_t^\tau = \int_{t-\tau}^t\sigma^2_s ds
$$
as integrated variance over the period $\tau$, and to
$$
RV_t^\tau = \sum_{i=0}^{n-1}\left(\log S_{t-\tau+(i+1)\delta}-\log S_{t-\tau+i\delta}\right)^2,
$$
where $\delta=\tau/n,\ n\in\mathbb{N}$, as realized variance, which is the simplest estimator of $IV_t^\tau$. 
It is worth noting that 
$$
\lim_{\tau\to 0} \frac{IV_t^\tau}{\tau}=\sigma^2_t,
$$
so fixing a small $\tau$ to obtain a good finite-difference approximation of $\sigma_t^2$, viewed as derivative of integrated variance, requires careful consideration on the accuracy with which $IV_t^\tau$ can be estimated using $RV_t^\tau$ (or similar alternative estimators; cf. \cite{ait2014high}). So far, we have discussed estimation topics for the mfOU process, without specifying a volatility measure for our model. In the rest of the paper, we will mainly focus on $Y_t=\log\sqrt{RV_t^\tau/\tau}$, with $\tau=1/252$ for daily volatility. This is common in the literature and has been used by \textcite{GJR18, WXY23, bibinger2025modeling}, among others. The daily horizon is the most natural choice for most applications and for specifying a global volatility system for stock markets, as trading-hour misalignments are less of an issue at this frequency.   

However, some applications might focus on volatility at shorter time horizons. In addition, since roughness is a local property of the process, \textcite{GJR18} pointed out the potential bias in estimating the Hurst coefficient from realized volatility $RV_t^\tau$ when the goal is to draw conclusions about the spot volatility $\sigma_t$, due to the smoothing effect of the integral operator. Other authors have approached the estimation problem by examining other proxies closer to spot volatility \parencite{BLP21, bolko2023gmm, todorov}. In order to support our modeling framework also for spot-volatility dynamics, we will provide numerical evidence for time scales up to $\tau=1/(252\times390)$ in Section \ref{increasing-frequency-sim} and we will fit the model to spot-volatility time series in Section \ref{intraday-volatility}. The latter application relies on spot-volatility time series obtained with the method introduced by \textcite{kristensen2010nonparametric}, which is a kernel-weighted version of $RV_t^\tau$ with $\tau$ serving as the kernel bandwidth. We employ a Gaussian kernel with bandwidth $\tau=5/3 \times 1/(252 \times 390)$, corresponding to 100 seconds.

\section{Monte Carlo study}\label{monte-carlo-study}

In this section, we evaluate the estimation procedure on synthetic data. 

\subsection{Simulation method}\label{simulation-method}

We can simulate the mfOU process using either exact or approximate methods. We primarily rely on the exact method of pre-multiplying the lower-triangular Cholesky factor of the covariance matrix obtained from Equation \eqref{eq:ccf} by a suitable matrix with standard Gaussian entries. We do so in Sections \ref{finite-sample-performance}, \ref{mc_dimensionality}, and \ref{slow-mean-reversion}. This methodology, while being exact, has the drawbacks that the expression in \eqref{eq:ccf} may be hard to compute for very large lags $k$, due to numerical issues in computing the integral, and the matrix inversion can be computationally demanding when long time series are required. The latter is the case  in Section \ref{increasing-frequency-sim}, where we mimic high-frequency observations. In this setting we therefore rely on an approximate method that combines a circulant embedding scheme for the driving fractional Gaussian noise with the Euler-Maruyama scheme for the mean-reverting dynamics. The former is an exact fast simulation method for stationary Gaussian processes that relies on the Fast Fourier Transform (FFT) and was initially introduced by \textcite{wood1994simulation} and later adapted by \textcite{ACLP10} to the mfBm. The Euler-Maruyama scheme, however, introduces a discretization error and starts the process from a non-stationary regime. We address the former problem by subsampling on finer partitions and the latter by discarding the initial 2/5ths of each trajectory. 
  
In each setting, we simulate discrete trajectories on the uniform partition of the interval $[0,T]$ with mesh $\Delta$. Given a set of parameters \((\alpha_i, H_i, \nu_i, \mu_i, \rho_{i,j}, \eta_{i,j}, \ i,j=1,\dots,N)\), we produce \(M\) trajectories of the mfOU process of length \(n=T/\Delta\),
$$
\left(Y_\Delta, Y_{2\Delta}, \dots, Y_{n\Delta}\right),
$$
where
$$
Y_{k\Delta}=\left(Y_{k\Delta}^1,\dots,Y_{k\Delta}^N\right),\ k=1,\dots,n.
$$
In Sections \ref{finite-sample-performance} and \ref{mc_dimensionality}, we set \(\mu_i=0,\ i=1,\dots,N, \ \Delta = 1/252,\ T=20\), and \(M=10^4\). The values of \(\Delta\) and \(T\) resemble the daily observations over roughly 20 years in the dataset employed in the main empirical analysis. Sections \ref{slow-mean-reversion} and \ref{increasing-frequency-sim} differ in that they use $M=10^3$. Section \ref{increasing-frequency-sim} also features smaller values of $\Delta$ to mimic intraday observations. The longer time series induced by the new partitions require the approximate simulation method, which is implemented by subsampling at intervals $\Delta$ from an even finer grid with mesh $\delta=\Delta/16$. Following \textcite{bolko2023gmm}, we use the GMM with the set of lags \(\mathcal{L}=(0,\ 1,\ 2,\ 3,\ 4,\ 5,\ 20,\ 50)\), which yields 31 moment conditions with 23 overidentifying restrictions in dimension 2. The rationale is that small-lag covariances are very informative about the parameters of the process, especially concerning its regularity. However, nearby lags  also exhibit high correlations. Therefore, larger-lag (20 and 50 days) covariances play a role as variance reduction device while also capturing longer-range properties.

\subsection{Finite sample performance}\label{finite-sample-performance}

We conduct Monte Carlo (MC) simulations in the bivariate case, $N = 2$, across nine different parameter sets to examine how the estimator behaves with varying values of \(\alpha_i, H_i,\ \rho_{i,j}\ \text{and}\ \eta_{i,j}\).

Table \ref{tab:mc1} presents the results. Each panel, labeled 1 through 9, reports the true assigned parameter values (True), the Bias, computed as difference between True and the average of point estimates across MC samples, and the MC sample standard deviation (Std Err).  \\
Panel 1 represents the baseline scenario, reflecting realistic values of the parameters estimated from the logarithm of realized volatility time series in the empirical analysis of Section \ref{empirical-analysis}. Here, we find reliable estimates for all parameters except for \(\alpha_1\ \text{and}\ \alpha_2\), which display slight upward biases and high standard errors - consistently with prior findings in the literature (\cite{andersen1996gmm, WXY23}). For the remaining parameters, standard errors remain low and biases are null.  
In Panel 2 and Panel 3 we let $\eta_{1,2}$ grow to $0.1$ and $0.2$, respectively, without any change in the quality of the estimates. Panel 4 and Panel 5 present results for varying $\rho_{1,2}$. In one case, we have independent components of the mfOU, while in the latter we have very correlated ones. In both cases the quality of the estimates remains unchanged.  Next, we look at changing Hurst coefficients. Panel 6 presents results for $H_1=0.1$ and $H_2=0.4$. In this case the standard errors of the univariate parameters pertaining to the second component and those of the correlation parameters, $\rho_{1,2}$ and $\eta_{1,2}$, appear higher. We expect the former effect to be related to the magnitude of $H_2$ while the second effect due to the overall $H_1+H_2$. Panel 7 shows a case of long memory, with $H_1 = 0.6$ and $H_2 = 0.7$, which is still covered by our asymptotic theory. Standard errors and biases are higher than the baseline scenario, possibly due to the fact that we approach a non-Gaussian situation (our conjecture for $max(H_1,H_2)>3/4$). Finally, let us focus on shrinking $\alpha_1$ and $\alpha_2$ in Panel 8 and Panel 9. As the speeds of mean reversion decrease, their biases grow, at least relatively to their true values. In these scenarios, standard errors overall increase and slight biases appear in some parameters.  
In general, the GMM estimator shows lower standard errors compared to the initial values given by the combination of \textcite{dgp1} and \textcite{WXY23}, as predicted by the theory, especially for \(\alpha_i,\ i=1,2\), \(\rho_{1,2}\), and \(\eta_{1,2}\).   

\begin{longtable}{@{}>{\raggedright\arraybackslash}p{2cm}l*{8}{c}@{}}
\caption{Finite sample performance of the GMM estimator on the simulated mfOU process. Biases and standard errors (Std Err) are computed as in-sample Monte Carlo quantities. $N = 2,\ M = 10^4,\ \Delta=1/252,\ T = 20,\ \mu_i=0,\ i=1,2$} \label{tab:mc1} \\
\toprule
 & & \textbf{$\alpha_1$} & \textbf{$\alpha_2$} & \textbf{$\nu_1$} & \textbf{$\nu_2$} & \textbf{$H_1$} & \textbf{$H_2$} & $\rho_{1,2}$ & $\eta_{1,2}$ \\
\toprule
\endfirsthead
\toprule
 & & \textbf{$\alpha_1$} & \textbf{$\alpha_2$} & \textbf{$\nu_1$} & \textbf{$\nu_2$} & \textbf{$H_1$} & \textbf{$H_2$} & $\rho_{1,2}$ & $\eta_{1,2}$ \\
\toprule
\endhead
\midrule
\multicolumn{9}{r}{\textit{Continued on next page}} \\
\endfoot
\bottomrule
\endlastfoot
\multirow{3}{*}{\textbf{Panel 1}} 
&\textbf{True} & \textbf{1.00} & \textbf{1.50} & \textbf{1.00} & \textbf{1.00} & \textbf{0.10} & \textbf{0.20} & \textbf{0.50} & \textbf{0.00} \\
&\textbf{Bias} & 0.13 & 0.14 & 0.00 & 0.01 & 0.00 & 0.00 & 0.00 & 0.00 \\
&\textbf{Std Err} &  0.53 & 0.63  & 0.04 & 0.08 & 0.01 & 0.02 & 0.06 & 0.03 \\
\midrule
\multirow{3}{*}{\textbf{Panel 2}} 
& \textbf{True} & 1.00 & 1.50 & 1.00 & 1.00 & 0.10 & 0.20 & 0.50 & \textbf{0.10} \\
& \textbf{Bias} &  0.14 & 0.13 & 0.00 & 0.01 & 0.00 & 0.00 & 0.00 & 0.00 \\
& \textbf{Std Err} & 0.54 & 0.59 & 0.04 & 0.07 & 0.01 & 0.02 & 0.06 & 0.03 \\
\midrule
\multirow{3}{*}{\textbf{Panel 3}} 
& \textbf{True} & 1.00 & 1.50 & 1.00 & 1.00 & 0.10 & 0.20 & 0.50 & \textbf{0.20} \\
&\textbf{Bias} & 0.14 & 0.13 & 0.00 & 0.01 & 0.00 & 0.00 & 0.00 & 0.00 \\
&\textbf{Std Err} & 0.54 & 0.56 & 0.04 & 0.07 & 0.01 & 0.02 & 0.05 & 0.03 \\
\midrule
\multirow{3}{*}{\textbf{Panel 4}} 
&\textbf{True} & 1.00 & 1.50 & 1.00 & 1.00 & 0.10 & 0.20 & \textbf{0.00} & 0.00 \\
&\textbf{Bias} & 0.08 & 0.13 & 0.00 & 0.01 & 0.00 & 0.00 & 0.00 & 0.00 \\
&\textbf{Std Err} & 0.48 & 0.58 & 0.04 & 0.07 & 0.01 & 0.02 & 0.10 & 0.04 \\
\midrule
\multirow{3}{*}{\textbf{Panel 5}} 
&\textbf{True} & 1.00 & 1.50 & 1.00 & 1.00 & 0.10 & 0.20 & \textbf{0.90} & 0.00 \\
&\textbf{Bias} & 0.11 & 0.13 & 0.00 & 0.00 & 0.00 & 0.00 & 0.00 & 0.00 \\
&\textbf{Std Err} & 0.49 & 0.59 & 0.04 & 0.07 & 0.01 & 0.02 & 0.01 & 0.01 \\
\midrule
\multirow{3}{*}{\textbf{Panel 6}} 
&\textbf{True} & 1.00 & 1.50 & 1.00 & 1.00 & \textbf{0.10} & \textbf{0.40} & 0.50 & 0.00 \\
&\textbf{Bias} & 0.13 & 0.28 & 0.00 & 0.05 & 0.00 & 0.02 & 0.00 & 0.01 \\
&\textbf{Std Err} & 0.53 & 0.77 & 0.04 & 0.20 & 0.01 & 0.07 & 0.06 & 0.07 \\
\midrule
\multirow{3}{*}{\textbf{Panel 7}} 
&\textbf{True} & 1.00 & 1.50 & 1.00 & 1.00 & \textbf{0.60} & \textbf{0.70} & 0.50 & 0.00 \\
&\textbf{Bias}      & 0.27 & 0.38 & 0.08 & 0.11 & 0.03 & 0.03 & -0.01 & 0.01 \\
&\textbf{Std Err}  & 0.47  & 0.57 & 0.16 & 0.19 & 0.07 & 0.07 & 0.20 & 0.27 \\
\midrule
\multirow{3}{*}{\textbf{Panel 8}} 
&\textbf{True} & \textbf{0.20} & \textbf{0.20} & 1.00 & 1.00 & 0.10 & 0.20 & 0.50 & 0.00 \\
&\textbf{Bias}      & 0.11 & 0.11 & 0.00 & - 0.01 & 0.00 & 0.01 & 0.02 & 0.00 \\
&\textbf{Std Err}  & 0.29 & 0.30 & 0.04 & 0.12 & 0.01 & 0.06 & 0.12 & 0.06 \\
\midrule
\multirow{3}{*}{\textbf{Panel 9}} 
&\textbf{True} & \textbf{0.05} & \textbf{0.05} & 1.00 & 1.00 & 0.10 & 0.20 & 0.50 & 0.00 \\
&\textbf{Bias}      & 0.11 & 0.17 & 0.00 & 0.01 & 0.00 & 0.06 & 0.02 & 0.01 \\
&\textbf{Std Err}  & 0.24 & 0.34  & 0.07 & 0.23 & 0.03 & 0.15 & 0.20 & 0.16 \\
\end{longtable}

\begin{figure}
    \centering
    \includegraphics[width=\linewidth]{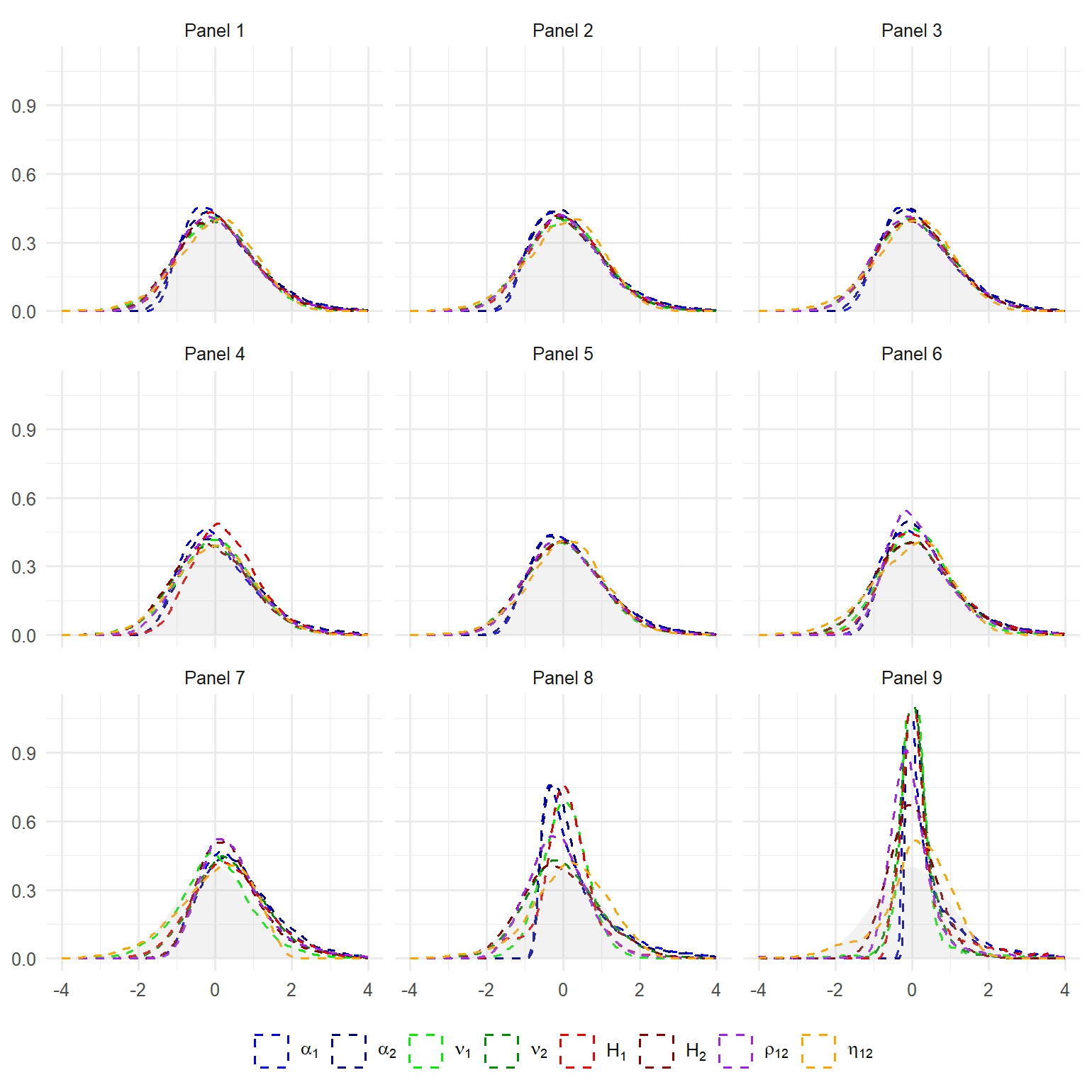}
    \caption{Kernel estimates of the densities of the elements in $(\hat\theta_n-\theta_0)/\widehat{s.e.}(\hat\theta_n)$, where $\widehat{s.e.}(\hat\theta_n)$ is the MC standard error of the GMM estimator $\hat\theta_n$.  Parameter settings are in Table \ref{tab:mc1}. Simulation parameters: $N = 2,\ M = 10^4,\ \Delta = 1/252,\ T = 20,\ \mu_{i}=0,\ i=1,2$.}
    \label{fig:mc1}
\end{figure}

Another perspective on the previous MC results is presented in Figure \ref{fig:mc1}, which shows the estimation error densities. This figure displays for each parameter setting, corresponding to the panels in Table \ref{tab:mc1}, the estimated error densities of all parameters, comparing the Nadaraya-Watson density of the standardized errors to a standard Gaussian distribution (shaded area). The colors of the dashed lines correspond to the different parameters as described in the legend.

From these results, we can conclude that a trajectory length of \(n=5000\approx20\times252\) is generally sufficient to approximate the normal distribution predicted by the asymptotic theory (Proposition \ref{theo:mde}). However, exceptions are observed in the distributions of the errors for \(\alpha_i,\ i=1,2\), which exhibit some skewness, especially for lower parameter values (Panel 8 and Panel 9). When \(\alpha_i,\ i=1,2\) are especially low, positive skewness and poorer fits to the normal distribution are also observed for the other parameters.

\subsection{Dimensionality}\label{mc_dimensionality}

\begin{figure}
    \centering
    \includegraphics[width=\linewidth]{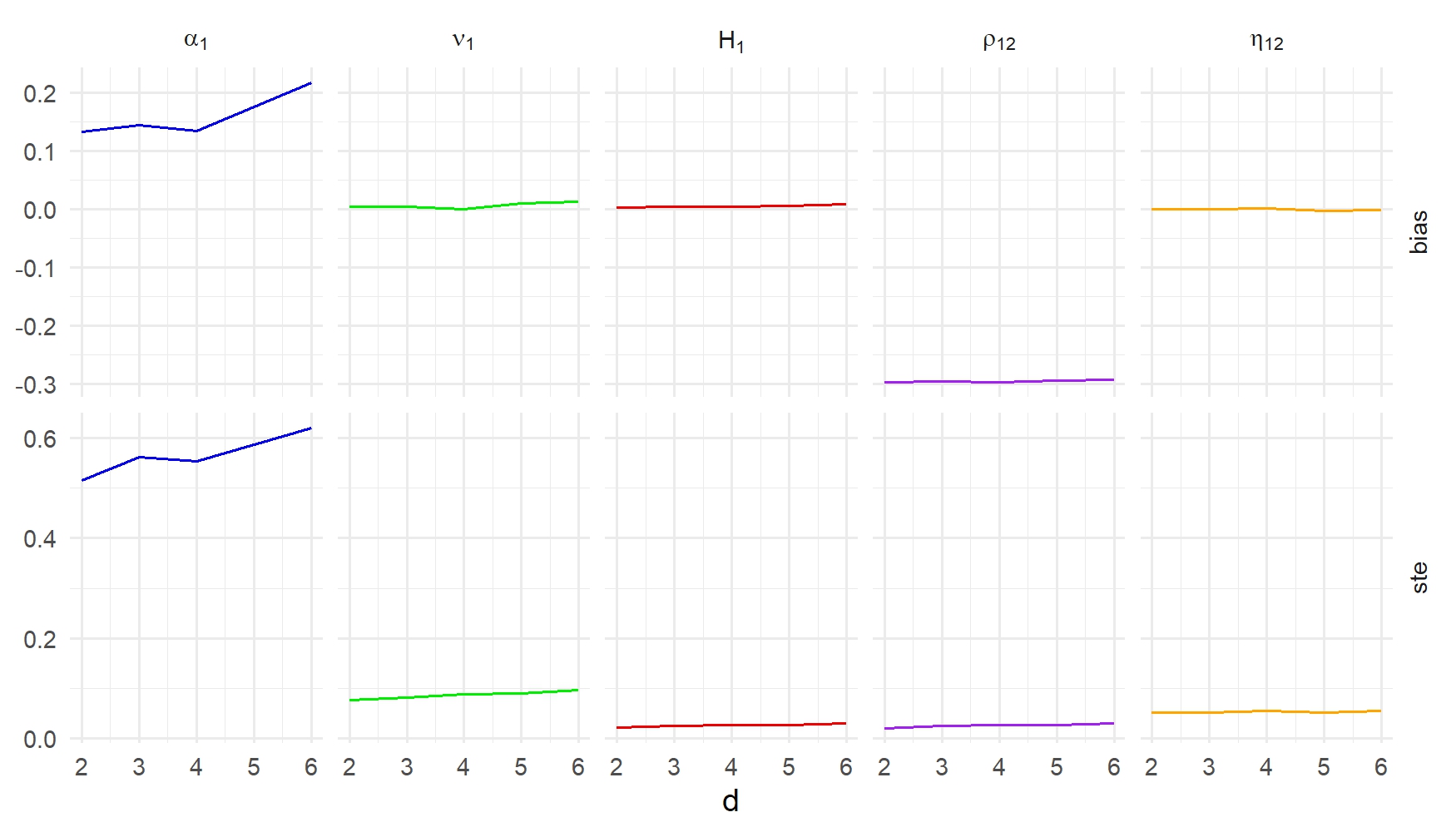}
    \caption{Bias and standard error of the GMM estimator as a function of the dimensionality of the mfOU process. Parameters are identical across components, $\alpha_i=1,\nu_i=1,H_i=0.1,\rho_{i,j}=0.5,\eta_{i,j}=0,\ i,j=1,\dots,N,\ M=10^3,\ \Delta = 1/252,\ T=20.$}
    \label{fig:mc3}
\end{figure}

We briefly consider how the estimation error changes with the dimensionality of the mfOU process. We simulate mfOU processes of dimension $N$ that grows from 2 to 6 and evaluate the biases and the standard errors in the univariate parameter estimates related to the first component, $\alpha_1\ ,\  \nu_1,\ H_1$, and the correlation parameters related to the first and second components, $\rho_{ 1,2}$ and $\eta_{1,2}$.  The parameters governing the mfOU processes are identical across the marginal components and fixed to the values for the first component in the baseline scenario presented in Table \ref{tab:mc1}. Pairs of components are ruled by $\rho_{i,j}=0.5,\ \eta_{i,j}=0\,\ i,j=1,\ \dots,N$, $N = 2,\ \dots, 6$. In all cases we consider $M=10^3$ trajectories of length $n=5000$. The results are shown in Figure \ref{fig:mc3}.

The bias in the speed of mean reversion seems to grow with the dimensionality of the process, whereas it is not clear whether the biases in the remaining parameters are growing or not. Standard errors clearly grow with the dimensionality of the process in all cases. A strong difference in the magnitude of both biases and standard errors is again apparent between the speed of mean reversion parameter and the remaining ones. 

\subsection{Slow mean reversion}\label{slow-mean-reversion}

\begin{figure}
    \centering
    \includegraphics[width=\linewidth]{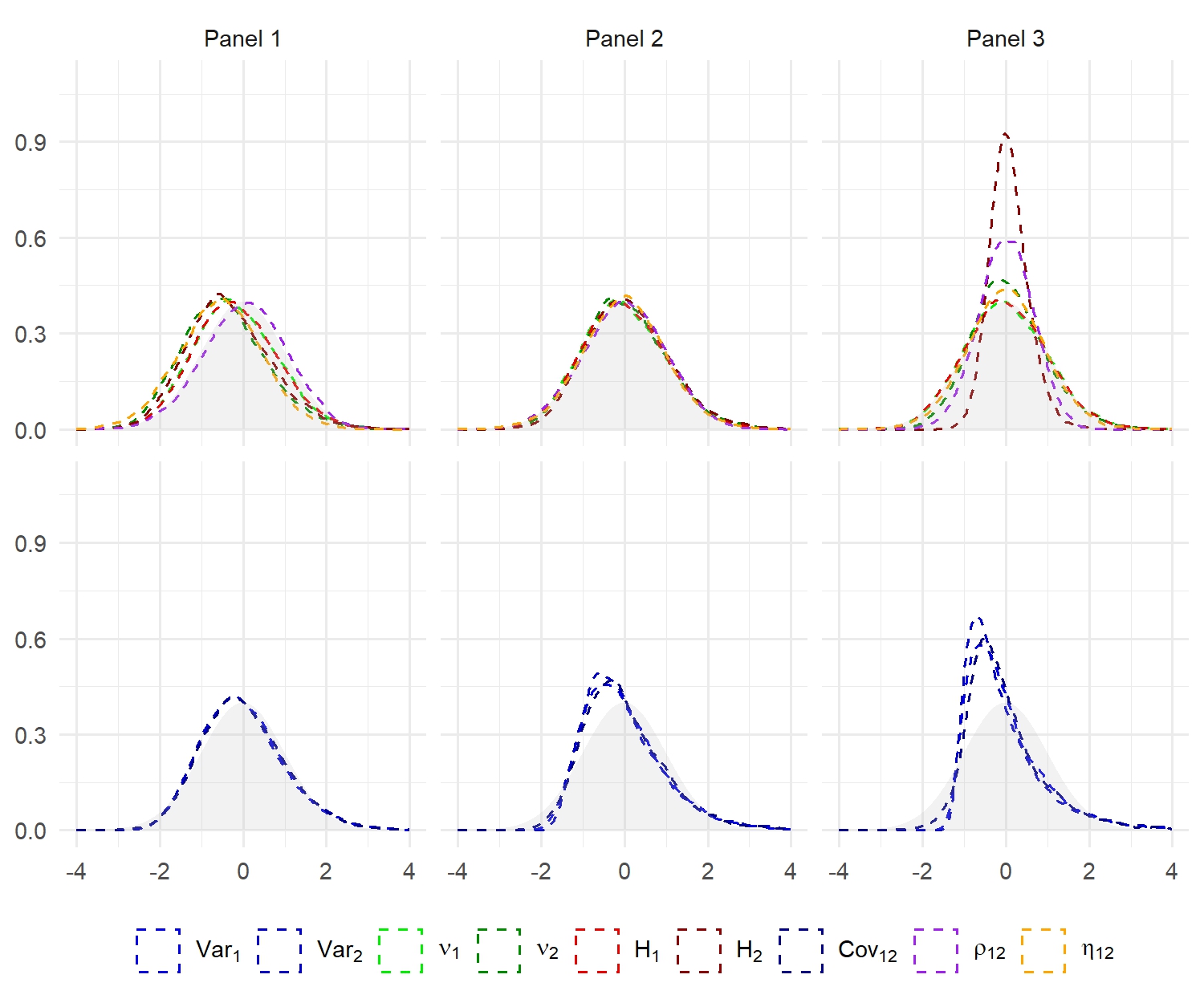}
    \caption{Kernel estimates of the densities of the elements in $(\hat\theta_n-\theta_0)/\widehat{s.e.}(\hat\theta_n)$, where $\hat\theta_n$ denotes the GMM estimator that uses asymptotic cross-covariance conditions, and $\widehat{s.e.}(\hat\theta_n)$ is the MC standard error of $\hat\theta_n$.  Parameter settings are in Table \ref{tab:mc2}. Simulation parameters: $N = 2,\ M = 10^4,\ \Delta = 1/252,\ T = 20,\ \mu_{i}=0,\ i=1,2$.}
    \label{fig:mc2}
\end{figure}

Motivated by the asymptotic expression of the cross-covariance function for \(\alpha\to 0\) given in Proposition \ref{eq:ccfl}, and the lower-quality results obtained for small \(\alpha_i,\ i=1,2\) in simulation in Section \ref{finite-sample-performance} and Section \ref{mc_dimensionality}, we also attempt GMM estimation using the small-alpha cross-covariance approximation from Proposition \ref{eq:ccfl} in the moment conditions. We include lag-$0$ variances and covariances (\(\text{V}_1,\ \text{V}_2,\ \text{C}_{1,2}\)), which appear in the asymptotic formula, as parameters to be estimated in the optimization process. The results are presented in Table \ref{tab:mc2} and Figure \ref{fig:mc2}.  

The results for \(\nu_i,\ H_i,\ i=1,2,\ \rho_{1,2},\ \text{and}\ \eta_{1,2}\) are satisfactory. We only observe slight biases in most parameters (Panel 1), which vanish as \(\alpha_i,\ i=1,2\) decrease (Panels 8 and 9). The standard errors are very similar to those obtained using the exact cross-covariance function \eqref{eq:ccf} in Table \ref{tab:mc1}. \\  
However, the same cannot be said for $\text{V}_1,\ \text{V}_2,$ and $\text{C}_{1,2}$, which exhibit a bias that originates similarly to the one previously associated with \(\alpha_i,\ i=1,2\). Motivated by this finding, we estimate the cross-covariance using a sample average estimator and observe similar evidence of biases across several lags when \(\alpha_i,\ i=1,2\) are small. This finding may explain the nature of the bias in the estimator of the mean reversion coefficient in \textcite{WXY23}, which relies on the sample variance.  

\begin{longtable}{@{}>{\raggedright\arraybackslash}p{2cm}l*{9}{c}@{}}
\caption{Finite sample performance of the GMM estimator based on asymptotic cross-covariance conditions on the simulated mfOU process. Biases and standard errors (Std Err) are computed as in-sample Monte Carlo quantities. $N = 2,\ M = 10^4,\ \Delta=1/252,\ T = 20,\ \mu_i=0,\ i=1,2$.} \label{tab:mc2} \\
\toprule
& & \textbf{$\nu_1$} & \textbf{$\nu_2$} & \textbf{$H_1$} & \textbf{$H_2$} & $\rho_{1,2}$ & $\eta_{1,2}$ & \textbf{$V_1$} & \textbf{$V_2$} & \textbf{$C_{1,2}$} \\
\toprule
\endfirsthead
\toprule
 & & \textbf{$\nu_1$} & \textbf{$\nu_2$} & \textbf{$H_1$} & \textbf{$H_2$} & $\rho_{1,2}$ & $\eta_{1,2}$ & \textbf{$V_1$} & \textbf{$V_2$} & \textbf{$C_{1,2}$} \\
\toprule
\endhead
\midrule
\multicolumn{9}{r}{\textit{Continued on next page}} \\
\endfoot
\bottomrule
\endlastfoot
\multirow{3}{*}{\textbf{Panel 1}} 
&\textbf{True} & \textbf{1.00} & \textbf{1.00} & \textbf{0.10} & \textbf{0.20} & \textbf{0.50} & \textbf{0.00} &  \textbf{0.46}  & \textbf{0.38} &  \textbf{0.21} \\
&\textbf{Bias} & -0.01 & -0.03 & 0.00 & -0.01 & 0.00 & -0.01 & 0.00 & 0.00 & 0.00 \\
&\textbf{Std Err} & 0.03 & 0.06 & 0.01 & 0.02 & 0.04 & 0.03 & 0.04 & 0.05 & 0.03 \\
\midrule
\multirow{3}{*}{\textbf{Panel 8}} 
& \textbf{True} & 1.00 & 1.00 & 0.10 & 0.20 & 0.50 & 0.00 &  \textbf{0.63}  & \textbf{0.84} &  \textbf{0.36} \\
& \textbf{Bias} & 0.00 & 0.00 & 0.00 & 0.00 & 0.00 & 0.00 & 0.00 & 0.00 & 0.00 \\
& \textbf{Std Err} & 0.03 & 0.06 & 0.01 & 0.02 & 0.04 & 0.03 & 0.12 & 0.27 & 0.03 \\
\midrule
\multirow{3}{*}{\textbf{Panel 9}} 
& \textbf{True} & 1.00 & 1.00 & 0.10 & 0.20 & 0.50 & 0.00 &  \textbf{0.84}  & \textbf{1.47} &  \textbf{0.55} \\
&\textbf{Bias} & 0.00 & 0.00 & 0.00 & 0.01 & 0.00 & 0.00 & 0.00 & -0.03 & -0.01 \\
&\textbf{Std Err} & 0.04 & 0.08 & 0.01 & 0.05 & 0.06 & 0.04 & 0.29 & 0.81 & 0.39 \\
\end{longtable}

Similar conclusions can be drawn from the estimation error densities shown in Figure \ref{fig:mc2}. They are slightly skewed in the baseline scenario (Panel 1) but quickly become centered and resemble a Gaussian distribution as the mean reversion coefficients decrease (Panel 8 and Panel 9). Minor exceptions are the densities of $H_2$ and $\rho_{1,2}$ which appear excessively peaked. In contrast, the variances (\(\text{V}_1,\ \text{V}_2\)) and the covariance (\(\text{C}_{1,2}\)) grow increasingly skewed and non-normal.

\clearpage
\subsection{Increasing frequency}\label{increasing-frequency-sim}

We now consider the case in which the time step between observations becomes smaller. While our main empirical results are based on daily observations ($\Delta=1/252$), some applications rely on intraday data. We therefore further divide the time step by $7$, $24$, $78$, and $390$, corresponding to hourly observations for markets open 7 or 24 hours per day, or to observations recorded every 5 minutes or every 1 minute for a market open 6.5 hours per day. These sampling frequencies are common in the modeling of financial variables, with higher frequencies being particularly relevant for studying spot volatility. As already noted, keeping $T = 20$ fixed while shrinking the time step expands the length of the time series, which necessitates approximate simulation techniques for generating synthetic data, as discussed in Section \ref{simulation-method}. 
We present the simulation results in Table \ref{tab:mc3}, where the parameters are set to the baseline scenario analyzed earlier. From these results, we conclude that the step size does not materially affect the performance of our estimator, which yields outcomes consistent with those obtained under the setting $\Delta = 1/252$ and the associated asymptotic theory. If anything, the last row of Table \ref{tab:mc3} suggests that performance may improve as more observations become available. This provides supportive evidence for using the mfOU process to model intraday dynamics.

\begin{longtable}{@{}>{\raggedright\arraybackslash}p{2.2cm} l *{8}{c} @{}}
\caption{Finite sample performance of the GMM estimator on the simulated mfOU processes for shrinking time step $\Delta$. 
Biases and standard errors (Std Err) are computed as in-sample Monte Carlo quantities.  $N = 2,\ M = 10^3,\ T = 20,\ \mu_i=0,\ i=1,2$}\label{tab:mc3} \\
\toprule
\textbf{$\Delta$} &  & \textbf{$\alpha_1$} & \textbf{$\alpha_2$} & \textbf{$\nu_1$} & \textbf{$\nu_2$} & \textbf{$H_1$} & \textbf{$H_2$} & $\rho_{1,2}$ & $\eta_{1,2}$ \\
\midrule
\endfirsthead
\toprule
\textbf{$\Delta$} &  & \textbf{$\alpha_1$} & \textbf{$\alpha_2$} & \textbf{$\nu_1$} & \textbf{$\nu_2$} & \textbf{$H_1$} & \textbf{$H_2$} & $\rho_{1,2}$ & $\eta_{1,2}$ \\
\midrule
\endhead
\midrule
\multicolumn{10}{r}{\textit{Continued on next page}} \\
\endfoot
\bottomrule
\endlastfoot
&  \textbf{True}
& \textbf{1.00} & \textbf{1.50} & \textbf{1.00} & \textbf{1.00} 
& \textbf{0.10} & \textbf{0.20} & \textbf{0.50} & \textbf{0.00} \\
\midrule
\multirow{2}{*}{$(\textbf{7}\times252)^{-1}$} 
& \textbf{Bias}    
& 0.17 & 0.19 & 0.00 & 0.01 & 0.00 & 0.00 & 0.00 & 0.00 \\
& \textbf{Std Err} 
& 0.56 & 0.62 & 0.03 & 0.05 & 0.00 & 0.02 & 0.06 & 0.02 \\
\midrule
\multirow{2}{*}{$(\textbf{24}\times252)^{-1}$} 
& \textbf{Bias}    
& 0.17 & 0.15 & 0.00 & 0.01 & 0.00 & 0.00 & 0.00 & 0.00 \\
& \textbf{Std Err} 
& 0.62 & 0.56 & 0.03 & 0.04 & 0.01 & 0.01 & 0.06 & 0.01 \\
\midrule
\multirow{2}{*}{$(\textbf{78}\times252)^{-1}$} 
& \textbf{Bias}    
& 0.18 & 0.17 & 0.00 & 0.00 & 0.00 & 0.00 & 0.00 & 0.00 \\
& \textbf{Std Err} 
& 0.57 & 0.55 & 0.02 & 0.03 & 0.01 & 0.02 & 0.06 & 0.03 \\
\midrule
\multirow{2}{*}{$(\textbf{390}\times252)^{-1}$} 
& \textbf{Bias}    
& 0.16 & 0.14 & 0.00 & 0.01 & 0.00 & 0.00 & 0.00 & 0.00 \\
& \textbf{Std Err} 
& 0.51 & 0.50 & 0.01 & 0.01 & 0.001 & 0.008 & 0.05 & 0.01 \\
\end{longtable}

\section{Fitting the model to empirical data}\label{empirical-analysis}

Our empirical study in Sections \ref{estimates} and \ref{slow-mean-reversion-1} utilizes realized volatility time series based on 5-minute price increments from the Realized Library of the Oxford-Man Institute.\footnote{\url{https://oxford-man.ox.ac.uk/research/realized-library} (no longer accessible)} The dataset we use spans over 20 years of daily observations, starting from January 3, 2000, to June 28, 2022. It includes realized volatility for 31 indices associated with stock exchanges worldwide. We retain only the series covering the entire sample period and discard the ones that have long sequences of missing values at the beginning, middle, or end of the sample. The final sample is composed of 22 time series with average length of 5,616 observations. We treat this collection as a multivariate system after removing observations with a realized volatility equal to zero, scaling to annual percentage points, applying the logarithm transformation and interpolating missing values with an autoregressive model of order five, AR(5). A detailed description of the dataset can be found in Appendix B. In Section \ref{intraday-volatility}, we employ a dataset of 1-minute prices for three American indices over the period from January 2015 to November 2025, obtained from FirstRate Data\footnote{\url{https://firstratedata.com/}}. These consist of around 390 observations per day, with a total of approximately 1,768,414 observations for each index. Once the spot-volatility time series are computed from these data, they are preprocessed as described above for realized volatility.

\subsection{Parameter estimates}\label{estimates}

Table \ref{tab:univ}, \ref{tab:rho}, and \ref{tab:eta} present the estimates obtained with the procedure described in Section \ref{estimation-method} on the Oxford-Man dataset. We employ the GMM estimation procedure with the same lags as for the MC study, i.e. \(\mathcal{L}=\left(0,\ 1,\ 2,\ 3,\ 4,\ 5,\ 20,\ 50\right)\), and measure time in years, \(\Delta=1/252\). Since \(L=\#\mathcal{L}=8\ \text{and}\ N=22\), we now have 528 parameters and 3641 moment conditions. 

Table \ref{tab:univ} presents the estimated parameters related to the univariate marginal components. The average log-volatility ranges from 2.14 for AORD to 2.76 for BVSP, with all of them being around 2.5. The mean reversion parameters, \(\alpha_i\), are mostly estimated around 0.6, with some exceptions. Most notably AORD, RUT, BVSP, and KSE exhibit higher values. 

Our estimates for the Hurst exponent \(H_i\) consistently fall below \(1/2\), averaging around 0.17 for the most liquid indices and a bit higher for the less liquid ones. They are slightly larger than what is usually found in the rough volatility literature.
The differences compared with other studies that focus on realized volatility data (\cite{GJR18, BLP21, WXY23}) can be explained by the fact that our estimates, obtained via GMM, are based on the decay of memory over the lags in \(\mathcal{L}=(0,\ 1,\ 2,\ 3,\ 4,\ 5,\ 20,\ 50)\), while other studies consider short-lag properties ($\Delta \to 0$), and by the specific period of our sample. The differences with studies that focus on spot volatility (e.g. \cite{BLP21, bolko2023gmm}) can be explained by the smoothing effect of the integral operator (see \cite{GJR18}).   
Recall now that \(H_i\) rules both the regularity of the trajectories and the memory of the process. Since our estimates take as input auto and cross-covariances with lags in $\mathcal{L}$, this suggests that the memory structure of the process over $50$ days is compatible with a multivariate model with rough trajectories. The model we are fitting will exhibit, as a consequence, absence of long memory at the univariate level and short-range interdependence at the multivariate level (cf. Section \ref{the-model}). However, since our estimates do not take in input large ($\ge 100$) lag auto and cross-covariances, we cannot conclude that empirical data do not show long range interdependence, since even at the univariate level roughness and long-memory can coexist. This is pointed out e.g. in \textcite{BLP21}, which would be a natural starting point for constructing a multivariate model allowing for both roughness and long-range interdependencies.

\begin{table}
    \centering
\caption{GMM estimates of the univariate marginal parameters on $\log\left(100\sqrt{RV\times252}\right)$, where $RV$ is the realized variance from the Oxford-Man library (rv5). The GMM procedure is based on $\mathcal{L}=\{0,\ 1,\ 2,\ 3,\ 4,\ 5,\ 20,\ 50\}$ and $\Delta=1/252$. Symbol is based on the provider convention.}
\renewcommand{\arraystretch}{1.2}
    \begin{multicols}{2} 
        \begin{tabular}{@{}>{\raggedright\arraybackslash}p{3cm}cccc@{}}
            \toprule
            \textbf{Symbol} & \(\mu\) & \(\alpha\) & \(\nu\) & \(H\) \\ 
            \midrule
            S5E      & 2.70 & 0.59 & 0.70 & 0.15 \\
            \cmidrule(r){1-1}
            SSMI     & 2.40 & 0.89 & 0.63 & 0.23 \\
            \cmidrule(r){1-1}
            IBEX     & 2.70 &  0.59 & 0.64 & 0.17 \\
            \cmidrule(r){1-1}
            GDAXI    & 2.68 & 0.40 & 0.64 & 0.17 \\
            \cmidrule(r){1-1}
            FTSE     & 2.54 & 0.62 & 0.65 & 0.18 \\
            \cmidrule(r){1-1}
            FCHI     & 2.63 & 0.50 & 0.64 & 0.17 \\
            \cmidrule(r){1-1}
            BFX      & 2.46 & 0.66 & 0.62 & 0.19 \\
            \cmidrule(r){1-1}
            AEX      & 2.53 & 0.56 & 0.67 & 0.18 \\
            \cmidrule(r){1-1}
            SSEC     & 2.67 & 0.64 & 0.64 & 0.26 \\
            \cmidrule(r){1-1}
            NSEI     & 2.53 & 0.84 & 0.73 & 0.19 \\
            \cmidrule(r){1-1}        
            N225     & 2.52 & 0.94 & 0.64 & 0.22  \\
            \bottomrule
        \end{tabular}
        
        \columnbreak 

        \begin{tabular}{@{}>{\raggedright\arraybackslash}p{3cm}cccc@{}} 
            \toprule
            \textbf{Symbol} & \(\mu\) & \(\alpha\) & \(\nu\) & \(H\) \\
            \midrule
            KSE      & 2.47 & 4.10 & 1.24 & 0.43 \\
            \cmidrule(r){1-1}
            KS11     & 2.54 & 0.52 & 0.56 & 0.32 \\
            \cmidrule(r){1-1}
            HSI      & 2.54 & 0.79 & 0.51 & 0.31 \\
            \cmidrule(r){1-1}
            BSESN    & 2.64 & 0.65 & 0.62 & 0.22 \\
            \cmidrule(r){1-1}
            AORD     & 2.14 & 1.58 & 0.72 & 0.28 \\
            \cmidrule(r){1-1}
            SPX      & 2.43 & 0.61 & 0.77 & 0.18 \\
            \cmidrule(r){1-1}
            RUT      & 2.35 & 1.49 & 0.79 & 0.17 \\
            \cmidrule(r){1-1}
            MXX      & 2.40 & 0.98 & 0.64 & 0.15 \\
            \cmidrule(r){1-1}
            IXIC     & 2.54 & 0.42 & 0.71 & 0.16 \\
            \cmidrule(r){1-1}
            DJI      & 2.44 & 0.64 & 0.74 & 0.18 \\
            \cmidrule(r){1-1}
            BVSP     & 2.76 & 1.44 & 0.61 & 0.19 \\
            \bottomrule
        \end{tabular}
    \end{multicols}
    \label{tab:univ}
\end{table}

\begin{figure}
    \centering 
    \includegraphics[width=0.79\linewidth]{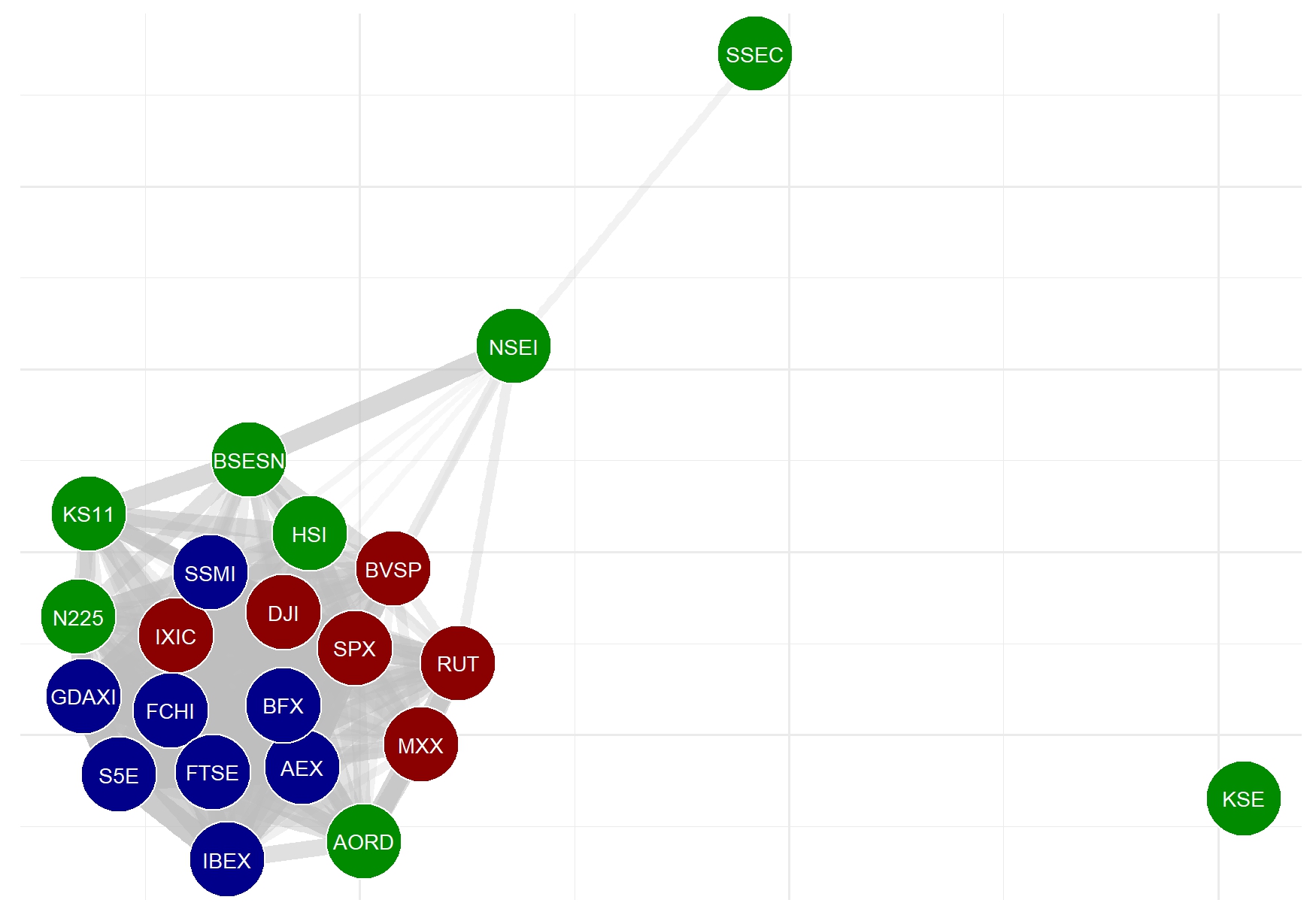} 
    \caption{Undirected graph representation of the estimates of $\rho_{i,j},\ i,j=1,\dots,N,\ i\ne j$. The nodes correspond to indices and the edges among pairs of indices have length inversely proportional to the GMM estimates of $\rho_{i,j}$ reported in Table \ref{tab:rho}.
    }
    \label{fig:rho} 
\end{figure}

Table \ref{tab:rho} and \ref{tab:eta} present the estimates of the matrices \(\rho\) and \(\eta\).  
Table \ref{tab:rho} exhibits the results for \(\rho\), the contemporaneous correlation coefficient of the underlying mfBm (\(\rho_{i,j}=\rho_{j,i}\)). The primary finding is that most values in \(\rho\) are high and positive, with only 33\% of the pairs showing values below 0.5. The strongest correlations are observed between the log-volatilities of SPX and DJI (0.98) and FCHI and FTSE (0.97). Several volatility pairs within Europe or North America display \(\rho_{i,j} > 0.7\), indicating that their underlying innovations tend to fluctuate together closely. In contrast, volatilities in Asia and Oceania (columns between SSEC and AORD) exhibit more moderate \(\rho_{i,j}\) values, with only two pairs, namely NSEI and BSESN and KS11 and BSESN, exceeding 0.7. The smallest coefficients for $\rho_{i,j}$ are observed between the volatilities for KSE and several others, or less notably between SSEC and others. In the former case they are closer to zero and once even negative. Similar conclusions emerge from the graphical representation of the estimates of \(\rho\) in Figure \ref{fig:rho}. The figure displays an undirected graph, obtained with the Fruchterman-Reingold algorithm (\cite{fruchterman1991graph}), where nodes correspond to indices, and edges are inversely proportional both in length and in width to the estimates of $\rho_{i,j}$. Edges corresponding to $\rho_{i,j}<0.4$ are deleted from the graph. This visualization shows how volatilities of North American and European indices cluster together more tightly, in contrast to the more dispersed patterns observed among Asian indices. 

In Table \ref{tab:eta} the estimated values of \(\eta\) are presented. Recall that \(\eta_{i,j}=-\eta_{j,i}\). Together with \(\alpha_i\) and \(\alpha_j\), \(\eta_{i,j}\) determines the degree of asymmetry in the cross-covariance function, which is directly related to the time irreversibility of the process.  More specifically, when \(H_{i,j}<1\) (short-range interdependence) both conditions \(\alpha_i > \alpha_j\) and \(\eta_{i,j} > 0\) result in a faster decrease of \(\gamma_{i,j}(k)\) compared with \(\gamma_{j,i}(k)\) as \(k\) increases. The opposite holds when the inequality is reversed for $H_{i,j}$ or $\alpha_i, \alpha_j$ and $\eta_{i,j}$. The absolute values in \(\eta\) range from $7\times 10^{-6}$, for the pair DJI and SPX, to 0.93 for KS11 and KSE. Two-thirds of the 231 pairwise relationships are characterized by \(|\eta_{i,j}|<0.05\), and \(90\%\) of them by \(|\eta_{i,j}|<0.1\). High absolute values in \(\eta\) are estimated for pairs in which the Asian indices KSE, SSEC, or KS11 are involved.
The Asian indices characterized by smaller absolute \(\eta_{i,j}\)s are BSESN and HSI. Among the lowest absolute values in \(\eta\), we observe $7\times 10^{-6}$ between SPX and DJI, $8\times 10^{-4}$ between IBEX and GDAXI, and $5\times 10^{-3}$ between FTSE and FCHI. These pairs are also characterized by similar \(\alpha_i\) and \(\alpha_j\) in Table \ref{tab:univ}, suggesting symmetric cross-covariances and time-reversibility. Generally, what we would consider to be interconnected exchanges, such as European and North American ones, show smaller absolute values in \(\eta\).\\    
The admissibility constraint ensuring the positive semidefiniteness of the covariance matrix (cf. Section \ref{the-model}) was not directly included in the optimization routine. However, it was satisfied by the estimates above.

\begin{landscape}
\begin{table}
    \centering
\scalebox{0.70}{
\begin{tabular}{rrrrrrrrrrrrrrrrrrrrrrrrrrrrrrrrrr}
    \toprule
 & S5E & SSMI & IBEX & GDAXI & FTSE & FCHI & BFX & AEX & SSEC & NSEI & N225 & KSE & KS11 & HSI & BSESN & AORD & SPX & RUT & MXX & IXIC & DJI & BVSP \\ 
  \midrule
S5E &  1& 0.89& 0.85&0.96 &0.90 &0.97 &0.90 &0.96&0.24 &0.33 &0.66 &0.04 &0.52 &0.56 &0.46 &0.62 &0.86 &0.62 &0.48 &0.73 &0.85 &0.62  \\ 
&  & & & & & & & & & & & & & & & & & & & & & \\
        SSMI     &  & 1 & 0.73 &0.900 &0.91 &0.90 &0.88 &0.94 &0.32 &0.44 &0.69 &0.07 &0.56 &0.61 &0.51 &0.67 &0.83 &0.62 &0.50 &0.70 &0.85 & 0.64 \\ 
&  &  & & & & & & & & & & & & & & & & & & & & \\ 
        IBEX     &  &  & 1 & 0.78 & 0.81& 0.88&0.85 &0.80 &0.14 &0.12 &0.53 &-0.06 &0.31 &0.48 &0.29 &0.65 &0.71&0.58&0.48&0.60&0.70&0.47       \\ 
&  &   & & & & & & & & & & & & & & & & & & & & \\ 
        GDAXI    &  &  &  & 1 & 0.87& 0.95& 0.88&0.95 &0.22 &0.30 &0.71 &0.08 &0.63 &0.62 &0.52 &0.56 &0.87 &0.53 &0.45 &0.79 &0.87 &0.65 \\ 
&  &   &  & & & & & & & & & & & & & & & & & & & \\ 
        FTSE     &  &  &  &  & 1 &0.93&0.92&0.92&0.31&0.39&0.66&0.09&0.57&0.67&0.52&0.76&0.88&0.73&0.66&0.76&0.88&0.65\\ 
&  &   &  &  & & & & & & & & & & & & & & & & & & \\ 
        FCHI     &  &  &  &  &  & 1 & 0.94 & 0.96& 0.24& 0.31& 0.68& 0.06& 0.56& 0.62& 0.49& 0.67& 0.88& 0.64& 0.55& 0.77& 0.88& 0.62\\ 
&  &   &  &  &  & & & & & & & & & & & & & & & & & \\ 
        BFX      & &  &  & &  &  & 1 & 0.93& 0.22& 0.33& 0.60& 0.07& 0.50& 0.61& 0.45& 0.72& 0.83& 0.68& 0.58& 0.72& 0.83& 0.58\\ 
&  &   &  &  &  &  & & & & & & & & & & & & & & & & \\ 
        AEX      &  &  &  &  & &  &  & 1 & 0.26& 0.36& 0.70& 0.08& 0.58& 0.64& 0.51& 0.66& 0.89& 0.64& 0.52& 0.78& 0.89& 0.66\\ 
&  &   &  &  &  &  &  & & & & & & & & & & & & & & & \\ 
        SSEC     &  &  &  &  &  &  &  &  & 1 & 0.46& 0.20& 0.05& 0.16& 0.35& 0.32& 0.36& 0.28& 0.29& 0.27& 0.13& 0.29& 0.33\\ 
&  &   &  &  &  &  &  &  & & & & & & & & & & & & & & \\ 
        NSEI     &  &  &  & & & & & & & 1 & 0.36& 0.19& 0.37& 0.40& 0.74& 0.38& 0.42& 0.51& 0.39& 0.24& 0.42& 0.54\\ 
&  &   &  &  &  &  &  &  &  & & & & & & & & & & & & & \\ 
        N225     &  &  &  &  &  &  &  &  &  &  & 1 & 0.10& 0.64& 0.64& 0.63& 0.46& 0.71& 0.40& 0.46& 0.71& 0.72& 0.64 \\ 
&  &   &  &  &  &  &  &  &  &  & & & & & & & & & & & & \\ 
        KSE      & &  & &  &  &  &  &  &  &  &  & 1 & 0.24& 0.06& 0.30& 0.01& 0.10& 0.01& 0.07& 0.15& 0.10& 0.06\\ 
&  &   &  &  &  &  &  &  &  &  &  & & & & & & & & & & & \\ 
        KS11     &  &  &  &  &  &  &  &  &  &  &  &  & 1 & 0.69& 0.72& 0.28& 0.68& 0.31& 0.41& 0.74& 0.69& 0.63 \\ 
&  &   &  &  &  &  &  &  &  &  &  &  & & & & & & & & & & \\ 
        HSI      &  &  &  &  &  & & & &  &  &  &  &  & 1 & 0.67& 0.55& 0.71& 0.54& 0.56& 0.73& 0.71& 0.61\\ 
&  &   &  &  &  &  &  &  &  &  &  &  &  & & & & & & & & & \\ 
        BSESN    &  &  & &  &  &  &  &  &  &  &  &  &  &  & 1 & 0.39& 0.61& 0.41& 0.52& 0.62& 0.60& 0.66 \\ 
&  &   &  &  &  &  &  &  &  &  &  &  &  &  & & & & & & & & \\ 

        AORD     &  &  &  &  &  &  &  &  &  &  & & &  &  &  & 1 & 0.64& 0.75& 0.69& 0.50& 0.63& 0.47\\ 
&  &   &  &  &  &  &  &  &  &  &  &  &  &  &  & & & & & & & \\ 
        SPX      &  &  &  &  &  &  &  &  &  &  &  &  &  &  &  &  & 1 & 0.73& 0.63& 0.91& 0.98& 0.68\\ 
&  &   &  &  &  &  &  &  &  &  &  &  &  &  &  &  & & & & & & \\ 
        RUT      &  &  &  &  &  &  &  &  &  &  &  &  &  &  &  &  &  & 1 & 0.70& 0.60& 0.71& 0.51\\ 
&  &   &  &  &  &  &  &  &  &  &  &  &  &  &  &  &  & & & & & \\ 
        MXX      &  &  &  &  &  &  &  &  &  &  &  &  & &  &  &  &  &  & 1 & 0.59& 0.62& 0.52\\ 
&  &   &  &  &  &  &  &  &  &  &  &  &  &  &  &  &  &  & & & & \\ 
        IXIC     &  &  &  &  &  &  &  &  &  &  &  &  &  &  &  &  &  &  &  & 1 & 0.89 & 0.63 \\ 
&  &   &  &  &  &  &  &  &  &  &  &  &  &  &  &  &  &  &  & & & \\ 
        DJI      &  &  &  &  &  &  &  & &  &  &  &  &  &  &  &  &  &  &  &  & 1 & 0.68 \\ 
&  &   &  &  &  &  &  &  &  &  &  &  &  &  &  &  &  &  &  &  & & \\ 
        BVSP     &  &  &  &  &  &  &  &  &  &  &  &  &  &  &  &  &  &  &  &  &  & 1 \\ 
&  &   &  &  &  &  &  &  &  &  &  &  &  &  &  &  &  &  &  &  &  & \\ 
\bottomrule
\end{tabular}
}
\caption{GMM estimates of $\rho_{i,j}$ on $\log\left(100\sqrt{RV\times252}\right)$, where $RV$ is the realized variance from the Oxford-Man library (rv5). The GMM procedure is based on $\mathcal{L}=\{0,\ 1,\ 2,\ 3,\ 4,\ 5,\ 20,\ 50\}$ and $\Delta=1/252$. Symbol is based on the provider convention.}
\label{tab:rho}
\end{table}
\end{landscape}

\begin{landscape}
\begin{table}
    \centering
\scalebox{0.70}{
\begin{tabular}{rrrrrrrrrrrrrrrrrrrrrrrrrrrrrrrrrr}
    \toprule
 & S5E & SSMI & IBEX & GDAXI & FTSE & FCHI & BFX & AEX & SSEC & NSEI & N225 & KSE & KS11 & HSI & BSESN & AORD & SPX & RUT & MXX & IXIC & DJI & BVSP \\ 
  \midrule
S5E &  0& 0& -0.01&-0.02 &-0.01 & -0.01 & 0.01 &-0.01&-0.07 &0.06 &0.05 &0.11 &-0.06 &0.02 &0.04 &0.06 &-0.03 &0.02 &-0.01 &-0.05 &-0.02 &0.04  \\ 
&  & & & & & & & & & & & & & & & & & & & & & \\
        SSMI     &  & 0 & -0.01 &-0.02 &-0.01 &-0.02 &0.03 &-0.01 &-0.12 &0.10 &0.04 &0.26 &-0.10 &0.04 &0.06 &0.10 &-0.05 &0.05 &0 &-0.08 & -0.04& 0.07 \\ 
&  &  & & & & & & & & & & & & & & & & & & & & \\ 
        IBEX     &  &  & 0 & 0 & -0.01 & -0.01 & 0.01 & 0 & -0.09 & 0.02 & 0.06 & -0.05 & -0.04 & 0.02 & 0.02 & 0.07 & -0.03 & 0.03 & -0.02 & -0.04 & -0.02 & 0.01       \\ 
&  &   & & & & & & & & & & & & & & & & & & & & \\ 
        GDAXI    &  &  &  & 0 & 0.01 & 0.00 & 0.02 & 0.01 & -0.06 & 0.06 & 0.08 & 0.20 & -0.03 & 0.06 & 0.05 &0.04 &0 & 0.01 &-0.01 &-0.04 &0 &0.08 \\ 
&  &   &  & & & & & & & & & & & & & & & & & & & \\ 
        FTSE     &  &  &  &  & 0 &0&0.02&0.01&-0.11&0.06&0.07&0.14&-0.05&0.07&0.04&0.12&-0.03&0.05&0.01&-0.05&-0.02&0.06\\ 
&  &   &  &  & & & & & & & & & & & & & & & & & & \\ 
        FCHI     &  &  &  &  &  & 0 & 0.02& 0.01& -0.09& 0.06& 0.08& 0.13& -0.06& 0.05& 0.04& 0.08& -0.01& 0.03& -0.01& -0.04& 0& 0.06\\ 
&  &   &  &  &  & & & & & & & & & & & & & & & & & \\ 
        BFX      & &  &  & &  &  & 0 & -0.02 & -0.12& 0.06& 0.04& 0.05& -0.05& 0.05& 0.05& 0.07& -0.04& 0.03& -0.01& -0.06& -0.04& 0.04\\ 
&  &   &  &  &  &  & & & & & & & & & & & & & & & & \\ 
        AEX      &  &  &  &  & &  &  & 0 & -0.08& 0.07& 0.06& 0.15& -0.07& 0.03& 0.05& 0.07& -0.02& 0.03& -0.01& -0.06& -0.02& 0.05\\ 
&  &   &  &  &  &  &  & & & & & & & & & & & & & & & \\ 
        SSEC     &  &  &  &  &  &  &  &  & 0 & 0.10& 0.12& 0.18& 0.16& 0.09& 0.16& 0.22& 0.11& 0.15& 0.14& 0.11& 0.11& 0.09\\ 
&  &   &  &  &  &  &  &  & & & & & & & & & & & & & & \\ 
        NSEI     &  &  &  & & & & & & & 0 & -0.02& 0.23& -0.11& -0.03& 0& 0.05& -0.06& 0.02& 0& -0.05& -0.06& -0.03\\ 
&  &   &  &  &  &  &  &  &  & & & & & & & & & & & & & \\ 
        N225     &  &  &  &  &  &  &  &  &  &  & 0 & 0.19& -0.24& -0.08& -0.05& -0.01& -0.10& -0.01& -0.05& -0.12& -0.09& -0.01\\ 
&  &   &  &  &  &  &  &  &  &  & & & & & & & & & & & & \\ 
        KSE      & &  & &  &  &  &  &  &  &  &  & 0 & -0.93& -0.49& -0.37& -0.11& -0.24& -0.08& -0.15& -0.24& -0.21& -0.32\\ 
&  &   &  &  &  &  &  &  &  &  &  & & & & & & & & & & & \\ 
        KS11     &  &  &  &  &  &  &  &  &  &  &  &  & 0 & 0.13& 0.11& 0.09& 0.08& 0.07& 0.07& 0.03& 0.08& 0.20 \\ 
&  &   &  &  &  &  &  &  &  &  &  &  & & & & & & & & & & \\ 
        HSI      &  &  &  &  &  & & & &  &  &  &  &  & 0 & 0.04& 0.06& -0.03& 0.02& 0.01& -0.06& -0.04& 0.12 \\ 
&  &   &  &  &  &  &  &  &  &  &  &  &  & & & & & & & & & \\ 
        BSESN    &  &  & &  &  &  &  &  &  &  &  &  &  &  & 0 & 0.03& -0.05& 0& 0& -0.07& -0.04& 0\\ 
&  &   &  &  &  &  &  &  &  &  &  &  &  &  & & & & & & & & \\ 
        AORD     &  &  &  &  &  &  &  &  &  &  & & &  &  &  & 0 & -0.10& -0.04& -0.08& -0.07& -0.10& -0.03 \\ 
&  &   &  &  &  &  &  &  &  &  &  &  &  &  &  & & & & & & & \\ 
        SPX      &  &  &  &  &  &  &  &  &  &  &  &  &  &  &  &  & 0 & 0.05& 0.01& -0.05& 0& 0.06\\ 
&  &   &  &  &  &  &  &  &  &  &  &  &  &  &  &  & & & & & & \\ 
        RUT      &  &  &  &  &  &  &  &  &  &  &  &  &  &  &  &  &  & 0 & -0.03& -0.04& -0.05& -0.03\\ 
&  &   &  &  &  &  &  &  &  &  &  &  &  &  &  &  &  & & & & & \\ 
        MXX      &  &  &  &  &  &  &  &  &  &  &  &  & &  &  &  &  &  & 0 & -0.01& -0.01& 0.02\\ 
&  &   &  &  &  &  &  &  &  &  &  &  &  &  &  &  &  &  & & & & \\ 
        IXIC     &  &  &  &  &  &  &  &  &  &  &  &  &  &  &  &  &  &  &  & 0 & 0.05& 0.07\\ 
&  &   &  &  &  &  &  &  &  &  &  &  &  &  &  &  &  &  &  & & & \\ 
        DJI      &  &  &  &  &  &  &  & &  &  &  &  &  &  &  &  &  &  &  &  & 0 & 0.07 \\ 
&  &   &  &  &  &  &  &  &  &  &  &  &  &  &  &  &  &  &  &  & & \\ 
        BVSP     &  &  &  &  &  &  &  &  &  &  &  &  &  &  &  &  &  &  &  &  &  & 0 \\ 
&  &   &  &  &  &  &  &  &  &  &  &  &  &  &  &  &  &  &  &  &  & \\ 
\bottomrule
\end{tabular}
}
\caption{GMM estimates of $\eta_{i,j}$ on $\log\left(100\sqrt{RV\times252}\right)$, where $RV$ is the realized variance from the Oxford-Man library (rv5). The GMM procedure is based on $\mathcal{L}=\{0,\ 1,\ 2,\ 3,\ 4,\ 5,\ 20,\ 50\}$ and $\Delta=1/252$. Symbol is based on the provider convention.}
\label{tab:eta}
\end{table}
\end{landscape}

\begin{figure}
    \centering 
    \includegraphics[width=\textwidth]{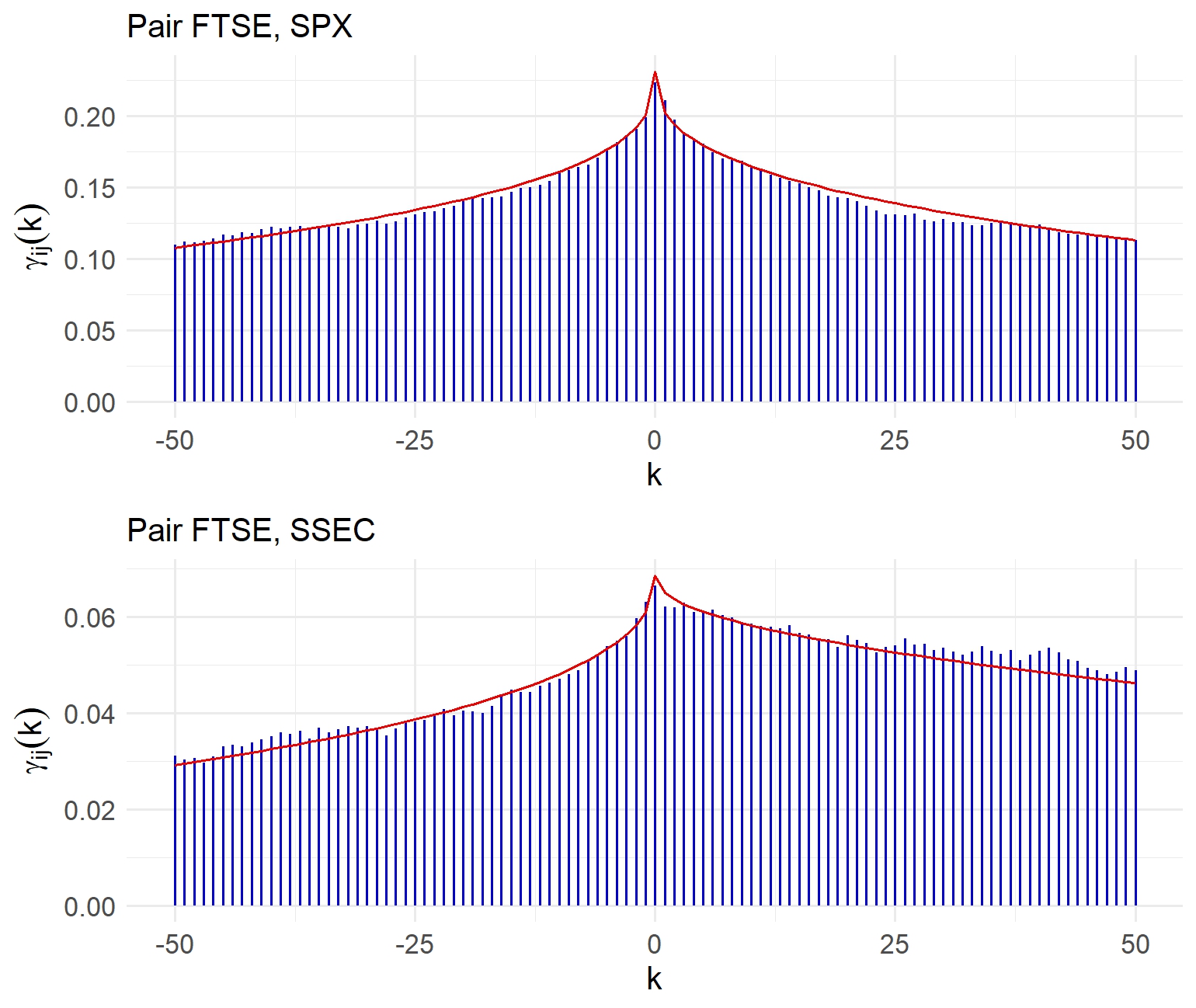} 
    \caption{Empirical cross-covariances of log-realized volatilities as blue bars, alongside the theoretical cross-covariances from our model, indicated by the red curves and based on the estimated parameters from Tables \ref{tab:univ}, \ref{tab:rho}, and \ref{tab:eta}. The upper panel shows the pair FTSE-SPX, while the lower panel shows the pair FTSE-SSEC.}
    \label{fig:ccfe} 
\end{figure}

To demonstrate the goodness of fit of our model and illustrate the degrees of asymmetry in the empirical cross-covariances of log-volatility, in Figure \ref{fig:ccfe} we present plots for the pairs FTSE-SPX and FTSE-SSEC. We identified these two extreme cases based on \(|\eta_{i,j}|\): when the value of the former is among the lowest ones and when it is among the highest ones. Notice that the asymmetry depends also on \(\alpha_i\) and \(\alpha_j\).

In the FTSE-SPX case, we have \(\eta_{i,j}=-0.02\), \(\alpha_{i}=0.62\), and \(\alpha_{j}=0.61\). The strong symmetry suggested by the close values of \(\alpha_i\) and \(\alpha_j\) and the small \(\eta_{i,j}\) is supported visually.  In the FTSE-SSEC case, where \(\eta_{i,j}=-0.11\), $\alpha_i=0.62$, and \(\alpha_j=0.64\), we expect asymmetry and indeed observe it. In the latter instance, since \(H_{i,j}<1\), both \(\eta_{i,j}<0\) and \(\alpha_i<\alpha_j\) accelerate the left-hand side of the cross-covariance towards zero.\footnote{The opposite holds when \(\eta_{i,j}>0\) and/or \(\alpha_i>\alpha_j\)} In the former case, $\eta_{i,j}<0$ and $\alpha_i>\alpha_j$ compensate each other. 

Our model fits well both these different behaviors. Moreover, as shown in Appendix \ref{d-empirics-of-marginal-components}, it effectively fits the autocovariance of the marginal components. The fit is also good for auto and cross-covariances of all the 22 components in the system, which overall exhibit characteristics closer to the symmetric case (see \href{https://ranieridugo.github.io/mfou_vol_appendix}{Online Appendix} for further empirical evidence).

\subsection{Slow mean reversion}\label{slow-mean-reversion-1}

\begin{figure}
    \centering 
    \includegraphics[width=\textwidth]{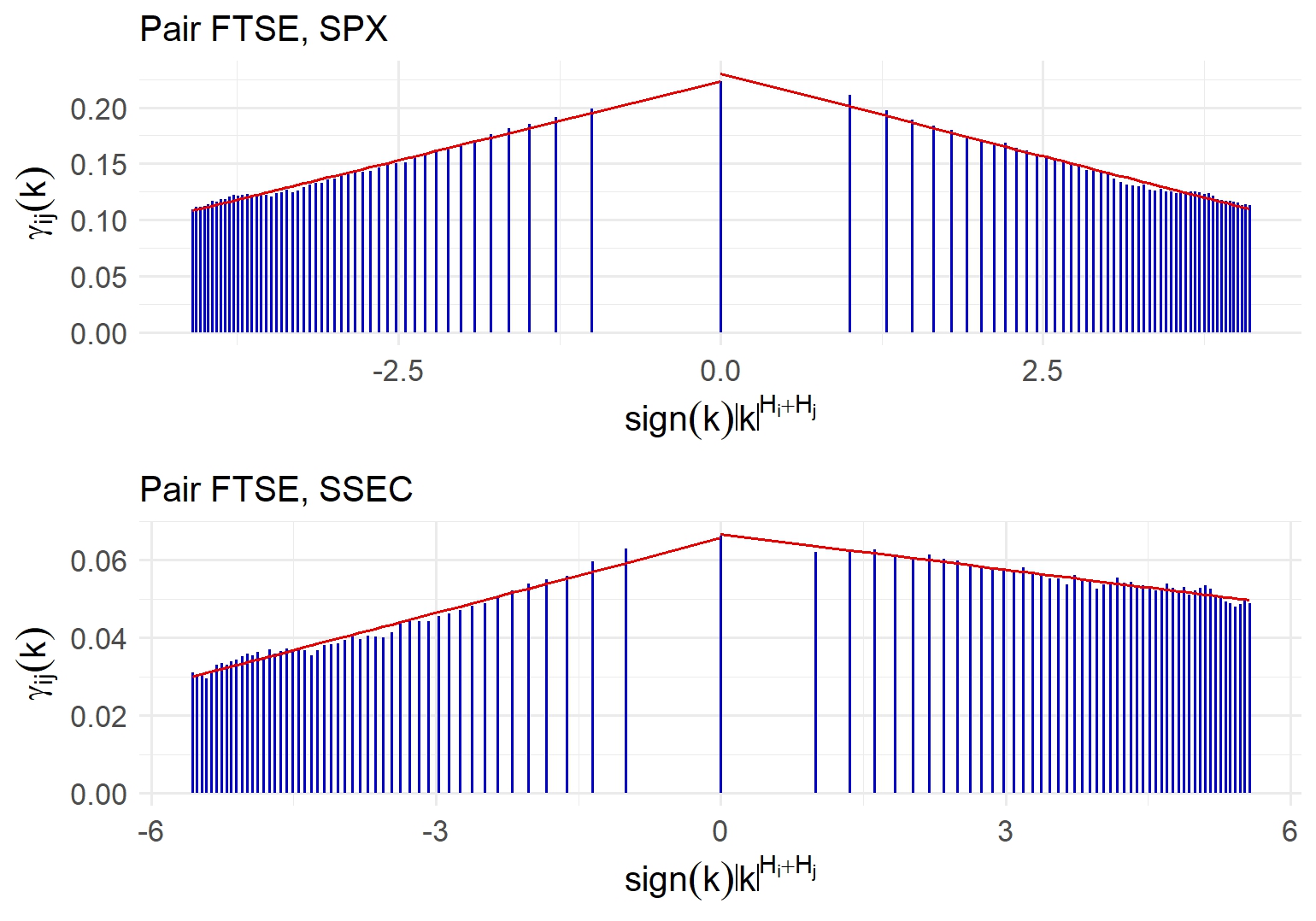}
    \caption{Empirical cross-covariances of log-realized volatilities as blue bars, plotted against a suitable power of the lag (given by the sum of the Hurst exponents), alongside the best linear fits, indicated by the red lines. The upper panel shows the pair FTSE-SPX, while the lower panel shows the pair FTSE-SSEC.}
    \label{fig:ccfl} 
\end{figure}

Following an approach similar to \textcite{GJR18}, we examine the empirical cross-covariance as a function of \(k^{H_i+H_j}\), where \(H_i\) and \(H_j\) are proxied by the estimated values in Table \ref{tab:univ}. According to the cross-covariance function of the mfOU process in the small \(\alpha\) (or small $k$) regime, this relationship should be approximately linear. Figure \ref{fig:ccfl} shows that this linear behavior holds very well in the data for lags up to 50 in both the symmtric (FTSE-SPX) and the asymmetric (FTSE-SSEC) cases. Evidence for autocovariances is provided in Appendix \ref{d-empirics-of-marginal-components}, with similar findings observed across the entire system available in the \href{https://ranieridugo.github.io/mfou_vol_appendix}{Online Appendix}.
Motivated by these findings, we attempt estimation of the system using the asymptotic cross-covariance function given in Proposition \ref{eq:ccfl} in the moment conditions of the GMM estimator. The resulting fit is strong, as demonstrated by Figure \ref{fig:ccfa}. Additional evidence on univariate marginals is provided in Appendix \ref{d-empirics-of-marginal-components}, with findings on the remaining series available in the \href{https://ranieridugo.github.io/mfou_vol_appendix}{Online Appendix}. Overall, parameter estimates for $H_i,\nu_i,\rho_{i,j}\ \text{and}\ \eta_{i,j},\ i,j=1,\dots,N$ are very similar between the two settings with the only difference that in the asymptotic setting, the parameters $\rho$ and $\eta$ appear lower in magnitude while the diffusion coefficients, $\nu_i,\ i=1,\dots,N$, appear higher (cf. \href{https://ranieridugo.github.io/mfou_vol_appendix}{Online Appendix}). The two effects compensate each other and result in very similar cross-covariance function evaluations in the two settings.
\begin{figure}
    \centering 
    \includegraphics[width=\textwidth]{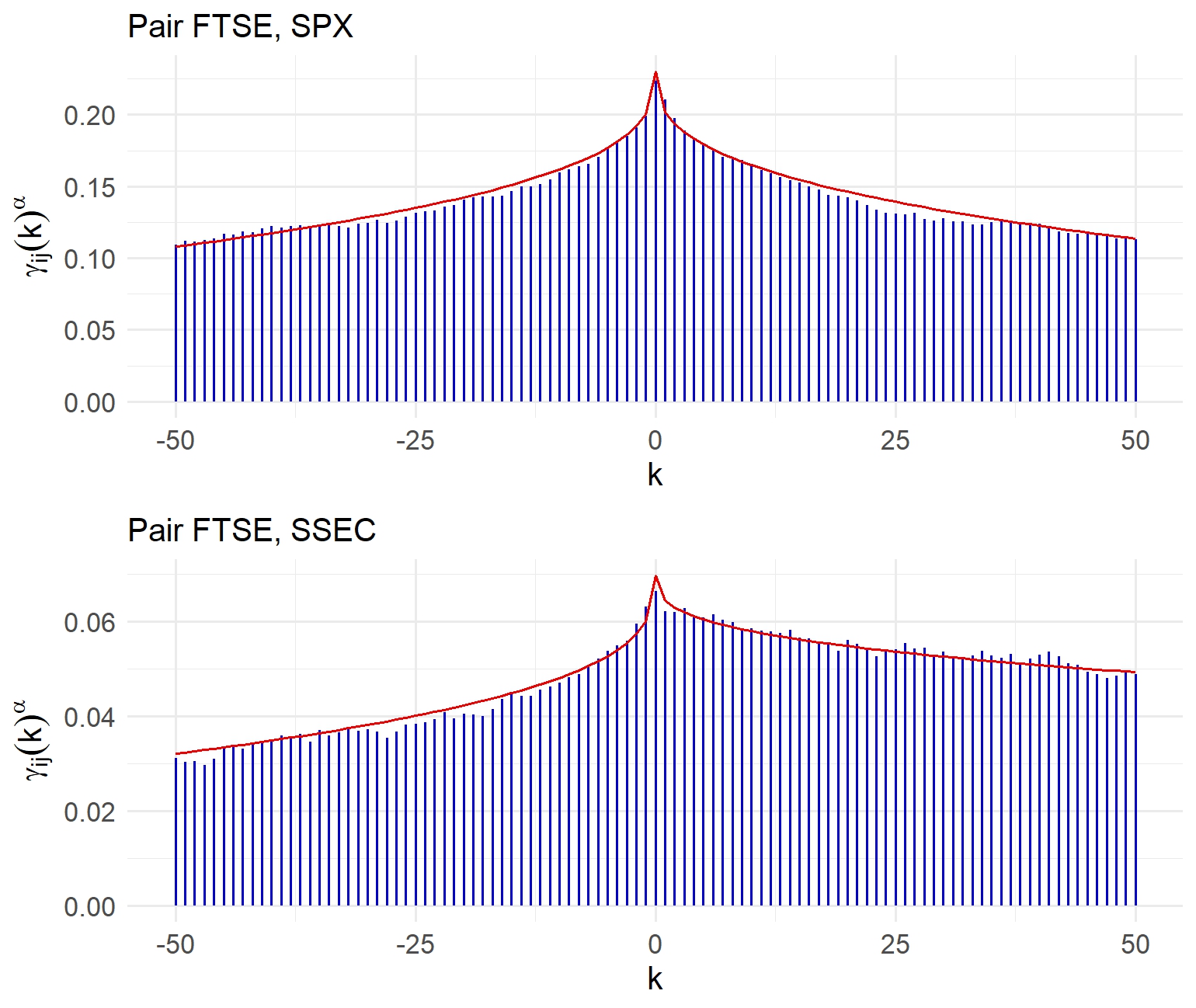}
    \caption{Empirical cross-covariances of log-realized volatilities as blue bars, alongside the approximate theoretical cross-covariances from our model for $\alpha\to 0$, indicated by the red curves and based on the estimated parameters (available in the \href{https://ranieridugo.github.io/mfou_vol_appendix}{Online Appendix}). The upper panel shows the pair FTSE-SPX, while the lower panel shows the pair FTSE-SSEC.}
    \label{fig:ccfa} 
\end{figure}
This can be interpreted as additional evidence for the ``near non-stationary regime'' of realized volatility found by \textcite{GJR18}.

\subsection{Intraday volatility}\label{intraday-volatility}
\begin{figure}
    \centering
    \includegraphics{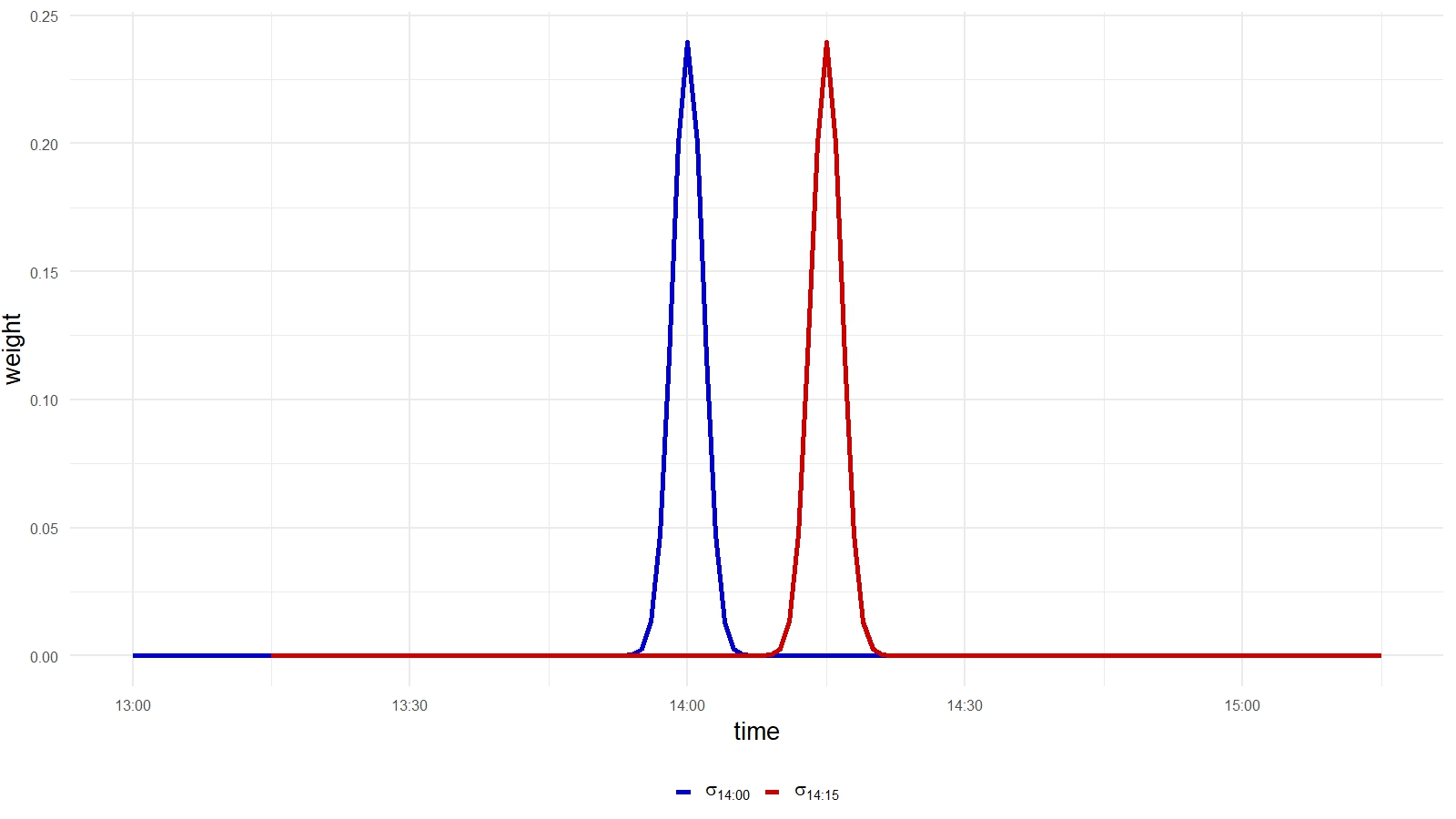}
    \caption{Weights used in the spot volatility estimator of \textcite{kristensen2010nonparametric}. Gaussian kernel, bandwidth equal to 100.}
    \label{fig:intraday}
\end{figure}

We now present the results from applying our model to intraday volatility dynamics. We use 1-minute prices for SPX, DJI, and IXIC for the period from January 2008 to November 2025 to estimate spot volatility using the kernel estimator proposed by \textcite{kristensen2010nonparametric}. The estimated quantity converges to the spot volatility as the bandwidth parameter goes to zero, similarly to what is discussed in Section \ref{volatility-measurement} for $IV_t^\tau/\tau$. We adopt a Gaussian kernel and fix the bandwidth to $\tau=5/3\times 1/(252\times390)$, corresponding to 100 seconds, in an attempt to incorporate as many returns as possible while minimizing overlapping returns in volatility estimates at a 15-minute distance. The overlapping reduces the degree of roughness. A graphical representation of the weighting scheme is presented in Figure \ref{fig:intraday} for volatility at 14:00 and 14:15 on a representative day. This estimator is not robust to microstructure noise, but we expect this quantity to be negligible for 1-minute prices of highly liquid assets.

\begin{figure}
    \centering 
    \includegraphics[width=\textwidth]{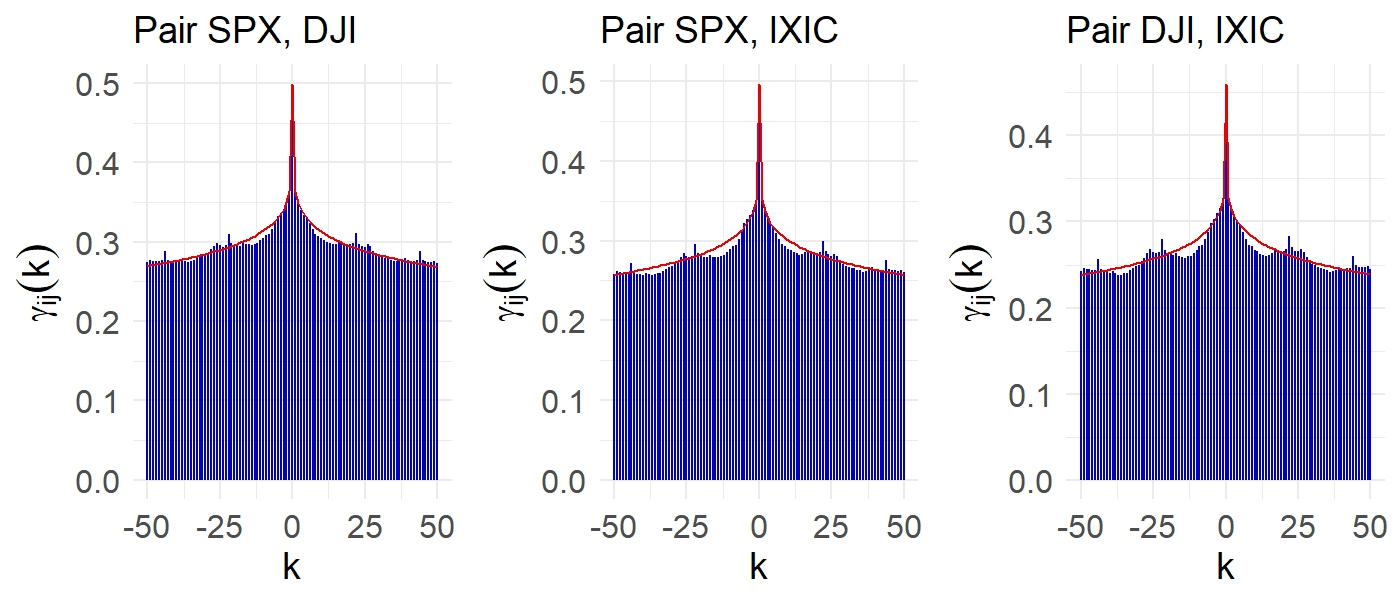} 
    \caption{Empirical cross-covariances of log-spot volatilities as blue bars, alongside the theoretical cross-covariances from our model, indicated by the red curves and based on the estimated parameters from Table \ref{tab:spot-est}. In this figure, the lag $k$ indicates the distance in terms of 15-minute intervals, $\Delta=1/(252\times 24)$.}
    \label{fig:ccfe_spot} 
\end{figure}

We address the intraday U-shaped seasonality by using the multiplicative decomposition
$$
\sigma_t=\sigma_t^s\tilde \sigma_t,
$$
where volatility $\sigma_t$ is decomposed into a deterministic seasonal component and a stochastic factor $\tilde \sigma_t$. To identify the seasonal component, we rely on the flexible Fourier form approach of \textcite{andersen1997intraday, andersen1998deutsche}. Both the spot-volatility estimation and the deseasonalization are implemented using the R package provided by \textcite{boudt2022analyzing}.
We retain the deseasonalized spot volatility at 15-minute intervals and scale it to annualized percentages using $100 \times \tilde\sigma_t \times \sqrt{252 \times 6.5 \times 60}$, where $\tilde\sigma_t$ is the kernel estimate of spot volatility at the minute scale. We then fit the mfOU process using the GMM estimator from Section \ref{estimation-method}, employing the same moment conditions as in Section \ref{estimates} and $\Delta = 1/(252 \times 24)$. The estimated parameters, which satisfy the admissibility constraint (cf. Section \ref{the-model}), are reported in Table \ref{tab:spot-est}, and a goodness-of-fit check based on cross-covariances is presented in Figure \ref{fig:ccfe_spot}.

\begin{table}[h!]
\centering
\caption{Estimated parameters of the mfOU process for SPX, DJI, and IXIC.}
\renewcommand{\arraystretch}{1.2} 
\begin{tabular}{@{}lccc|lccc@{}}
\toprule
\textbf{} & \textbf{SPX} & \textbf{DJI} & \textbf{IXIC} &
\textbf{} & \textbf{SPX-DJI} & \textbf{SPX-IXIC} & \textbf{DJI-IXIC} \\
\midrule
$\mu$ & 2.74 & 2.72 & 2.78 & $\rho$ & 0.98& 0.96 & 0.93 \\
$\alpha$ & 0.21 & 0.30 & 0.30 & $\eta$ & 0 & 0 & 0 \\
$\nu$ & 0.96 & 0.94 & 0.96 &  &  &  &  \\
$H$ & 0.07 & 0.07 & 0.06 &  &  &  &  \\
\bottomrule
\end{tabular}
\label{tab:spot-est}
\end{table}

Comparing Table \ref{tab:spot-est} with Tables \ref{tab:univ}, \ref{tab:rho}, and \ref{tab:eta}, we conclude that spot and integrated quantities exhibit similar average levels as well as similar multivariate parameters. However, spot volatility features slightly different mean-reversion and diffusion parameters and, most notably, smaller Hurst coefficients. The smaller Hurst coefficients in spot volatility have been already documented by \textcite{bolko2023gmm} and \textcite{BLP21} and can be explained by the smoothing effect of the integral operator \parencite{GJR18}. Spot volatility displays higher covariances than realized volatility, as shown in Figure \ref{fig:ccfe_spot}. Moreover, the decay of the cross-covariance follows a slightly different pattern, which is still accurately captured by our model, with the exception of small discrepancies caused by the residual seasonality. Additional evidence for univariate marginals is available in Appendix \ref{d-empirics-of-marginal-components}. This evidence suggests that our model can be used to capture spot-volatility dynamics. However, working with spot volatility is more challenging due to the intraday seasonality, the availability of price data for its estimation, and, in particular, the lack of aligned trading hours across markets. We therefore present these results as an illustration of the model’s potential applications and will focus on realized volatility for our spillover analysis.

\section{Spillover analysis}\label{spillovers}

Prompted by the asymmetries in the empirical cross-covariances observed among some pairs of log-realized volatilities, we study the effects of these lead-lag relationships in the framework of spillovers (\cite{diebold2009measuring, diebold2012better}). Spillovers are a method for quantifying the lead-lag relationships in a time series system exploiting the concept of forecast error variance decomposition, which was first introduced in \textcite{pesaran1997working, pesaran1998generalized}. The forecast error variance decomposition determines how much variability in forecasting a component of the system, i.e. calculating its future expectation, is attributable to the innovations in another component of the system occurring during the forecast horizon. This idea encapsulates the delayed influence of the dynamics of one component over another and corresponds to the building block for spillover measures, which is also denoted directional pairwise spillover. Aggregating these quantities appropriately, one can recover the influence exerted among groups of components.  
Let \(\psi_{i,j}(h)\), be the share of the variance in the error of predicting \(Y^i_{t+h}\) due to innovations in the variable \(\left(Y^j_{s}\right)_{s\in[t,t+h]}\), for \(i,j=1,\dots,N\). In \textcite{pesaran1997working, pesaran1998generalized}, in the discrete time setting, this is defined as
\begin{equation*}
\psi_{i,j}(h)=
\frac{\mathbb{E}\left[\left(Y_{t+h}^i-\mathbb{E}\left[Y_{t+h}^i|\mathcal{F}_{t-1}\right]\right)^2\right]-
\mathbb{E}\left[\left(Y_{t+h}^i-\mathbb{E}\left[Y_{t+h}^i|\mathcal{F}_{t-1},\left(\varepsilon^j_{t+k}\right)_{k=0}^{h}\right]\right)^2\right]}
{\mathbb{E}\left[\left(Y_{t+h}^i-\mathbb{E}\left[Y_{t+h}^i|\mathcal{F}_{t-1}\right]\right)^2\right]},
\end{equation*}
where \(\varepsilon_{t}\) is the $N$-variate white noise innovation in \(Y_t\) at time $t$, \(\varepsilon^j_{t}\) is its $j$-th component, and \(\mathcal{F}_{t-1}\) is the information set at time \(t-1\), i.e. the filtration generated by the underlying \(\varepsilon_{t-\ell},\ \ell=1,\dots\). In order to use the information in the variance decomposition matrix \(\Psi\), with entries \(\left[\Psi(h)\right]_{i,j}=\psi_{i,j}(h)\), to construct spillover indices in the presence of cross-sectionally correlated white noises, \textcite{diebold2012better} consider the normalized quantities 
\begin{equation}
\label{eq:psi}
\widetilde{\psi}_{i,j}(h)=\frac{\psi_{i,j}(h)}{\sum_{j=1}^N\psi_{i,j}(h)}.
\end{equation}
In order to carry out the spillover analysis, we manipulate our model to get as close as possible to the framework of \textcite{diebold2012better}, that is by considering the causal and discretized version of the mfOU process represented in terms of innovations uncorrelated over time. The mfOU process admits an integral moving average representation where the innovations are white noise (cf. \cite{dgp1} and Appendix \ref{e-spillovers-methodology}). Additionally, since we are interested in the effect of past innovations on future dynamics, we restrict our attention to the causal version of the mfOU process, that is, when the process depends only on past values of the driving white noise (see \cite{ACLP10} and  Appendix \ref{e-spillovers-methodology} for details). In this setting, we lose one degree of freedom in the choice of the parameters, which we can see as fixing \(\eta_{i,j}=f(H_i,H_j,\rho_{i,j})\) for a function \(f\) given in \textcite{ACLP10}. This constraint implies that \(\eta_{i,j}=0\) if and only if \(H_i=H_j\). With these assumptions, we are able to derive a closed form expression for the quantity in (\ref{eq:psi}) within our model.

\begin{prop}
In the adapted time discretization of the causal mfOU and mfBM processes (see Appendix \ref{e-spillovers-methodology}), assuming $t>h$, we have
$$
\widetilde\psi_{i,j}(h)=\frac{G^2_{i,j}/\sqrt{G_{j,j}}}{\sum_{k=1}^N G_{i,k}^2/\sqrt{ G_{k,k}}},
$$
where 
$$
 G_{i,j}=\sqrt{\frac{B(H_i+\frac{1}{2},H_i+\frac{1}{2})B(H_j+\frac{1}{2},H_j+\frac{1}{2})}{\sin\left(\pi H_i\right)\sin\left(\pi H_j\right)}}\frac{1}{B(H_i+\frac{1}{2},H_j+\frac{1}{2})}\frac{\sin\left(\pi(H_i+H_j)\right)}{\cos\left(\pi H_i\right)+\cos\left(\pi H_j\right)}\rho_{i,j},
$$
and $B(x,y),\ x,y>0$, denotes the Beta function.
\end{prop}

The proof is given in Appendix E.

\begin{remark} In the adapted discretization of the moving average representation of the causal mfOU process, the forecast error variance shares, $\widetilde\psi_{i,j}$, and the resulting spillover indices are independent of the forecast horizon, $h$, the history of the process, $\mathcal{F}_{t-1}$, and the discretization step. 
\end{remark}

We proceed to estimate the causal model following the same methodology detailed in Section \ref{estimation-method} with the additional constraint \(\eta_{i,j}=f\left(H_i,H_j,\rho_{i,j}\right)\), calculate \(\widetilde\psi_{i,j}\) with the estimated parameters, and construct the spillover indices introduced by \textcite{diebold2012better}, which are summarized in Table \ref{tab:spill}. Note that the additional constraint does not affect the asymptotic theory of the estimator. Estimates obtained in the causal setting are available in the \href{https://ranieridugo.github.io/mfou_vol_appendix}{Online Appendix}.

\begin{table}
\centering
\caption{Definitions of Spillover Indices from \textcite{diebold2012better}, where $\widetilde\psi_{i,j}(h)$ is defined in (\ref{eq:psi}).}
\begin{tabular}{@{}lp{6cm}l@{}}
\toprule
\textbf{Spillover Index} & \textbf{Description} & \textbf{Definition} \\ \midrule
Total & 
Aggregation over all components in the system of the normalized forecast error variance due to shocks in other components.
&
$S(h) = \frac{\sum_{i,j=1, i \neq j}^N \widetilde{\psi}_{ij}(h)}{N} \cdot 100$ \\[10pt] 
\cmidrule(r){1-3}
Received & 
Share of total spillovers received by a component of the system from all the other components. &
$S_{i,\cdot}(h) = \frac{\sum_{j=1, j \neq i}^N \widetilde{\psi}_{ij}(h)}{N} \cdot 100$ \\[10pt] 
\cmidrule(r){1-3}
Transmitted & 
Share of total spillovers transmitted by a component of the system to all the other components. &
$S_{\cdot,i}(h) = \frac{\sum_{j=1, j \neq i}^N \widetilde{\psi}_{ji}(h)}{N} \cdot 100$ \\[10pt] 
\cmidrule(r){1-3}
Net & 
Difference between the spillovers transmitted those received for each component. &
$S_{i}(h) = S_{\cdot,i}(h) - S_{i,\cdot}(h)$ \\[10pt] 
\cmidrule(r){1-3}
Net Pairwise & 
Difference between the spillovers transmitted from component \(i\) to component \(j\) and those transmitted from \(j\) to \(i\). &
$S_{i,j}(h) = \left( \frac{\widetilde{\psi}_{ji}(h) - \widetilde{\psi}_{ij}(h)}{N} \right) \cdot 100$ \\ 
\bottomrule
\end{tabular}
\label{tab:spill}
\end{table}

\begin{figure}
    \centering 
    \includegraphics[width=\textwidth]{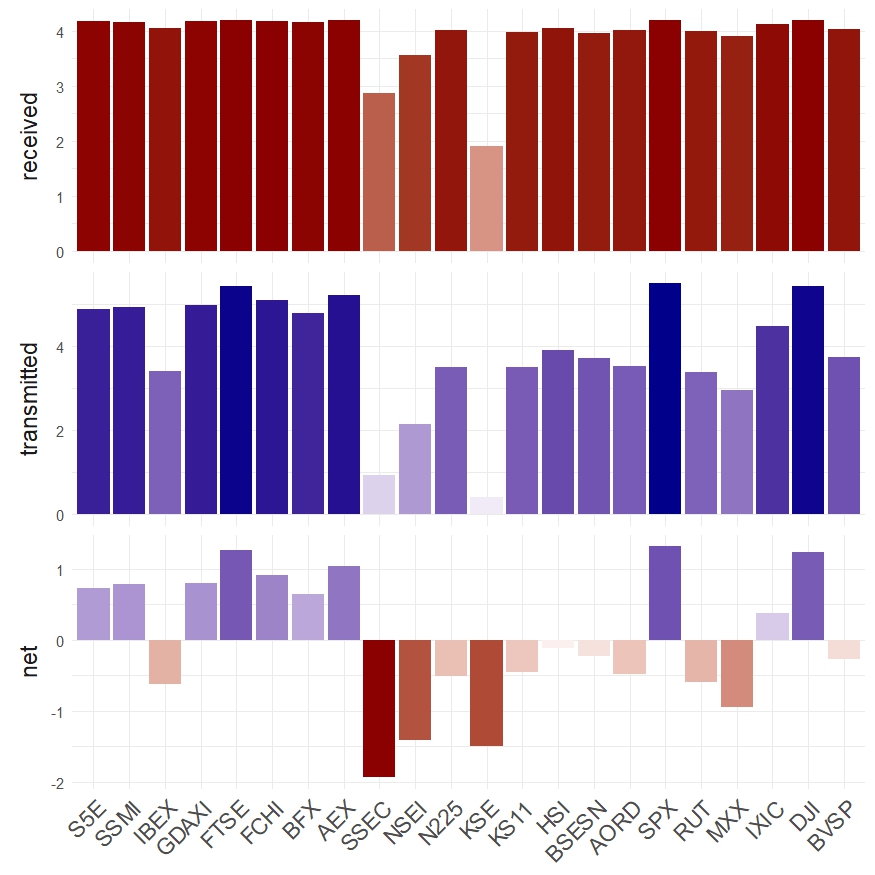} 
    \caption{Estimates of directional volatility spillovers over the whole period (2000-2022) for each index, obtained with the parameters estimated on log-realized volatility time series. The figures in the first and second rows represent aggregated shares of the forecast error variance decomposition. See definitions in Table \ref{tab:spill}.}
    \label{fig:spillovers_directional} 
\end{figure}

\begin{figure}
    \centering 
    \includegraphics[width=\textwidth]{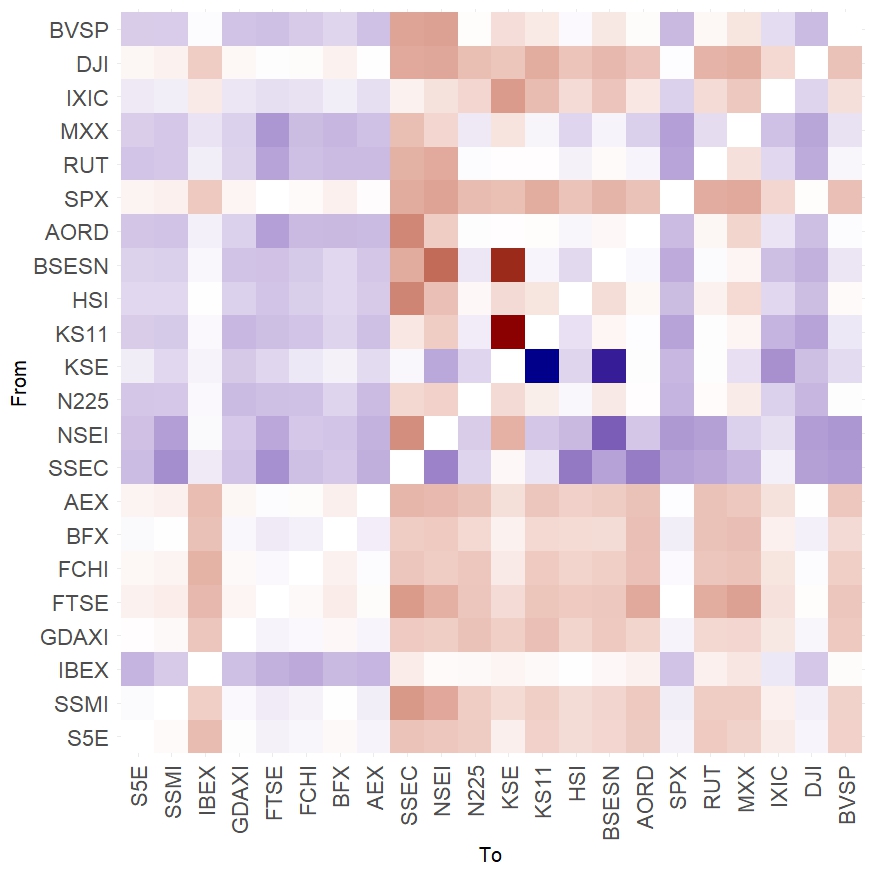} 
    \caption{Estimates of net pairwise volatility spillovers over the whole period (2000-2022) for each combination of indices i,j, obtained with the parameters estimated on log-realized volatility time series. Each entry is the difference between the volatility shocks transmitted from index i to index j and those transmitted from j to i. See definition in Table \ref{tab:spill}.}
    \label{fig:spillovers_pairwise} 
\end{figure}

We obtain the following results over the whole period from January 2000 to June 2022.
\begin{itemize}
\item
  \textit{Total spillovers} amount to 86\% of the normalized forecast error variance decomposition of log-realized volatilities.
\item
  \textit{Directional spillovers} are illustrated in Figure \ref{fig:spillovers_directional}. The top panel shows that the amount of received spillovers is quite uniform among European and North American volatilities, with greater differences among other components. Their level of variability is a direct consequence of the normalization step in (\ref{eq:psi}). Greater variability is preserved in the spillovers transmitted to others, in the middle panel, resulting in distinguishable net quantities. In fact, as shown in the bottom panel, European and North American indices generally transmit at least as much volatility spillover as they receive, with SPX, DJI, and FTSE playing major roles as transmitters, while Asian indices consistently exhibit negative net volatility spillovers. An exception in Europe is IBEX, which shows negative net volatility spillovers. The KSE index appears to be the most isolated in our sample, with relatively low levels of both transmitted and received volatility spillovers.
\item
 \textit{Net pairwise spillovers}, presented in Figure \ref{fig:spillovers_pairwise}, offer a more granular perspective. Although this measure cannot be interpreted in terms of shares of the normalized forecast error variance, net pairwise spillovers provide valuable qualitative insights. Notably, FTSE is the only index with all positive net pairwise volatility spillovers, followed by SPX, which only exhibits negative net volatility spillovers against FTSE. Other primary net transmitters include DJI and AEX. The main net receivers appear to be SSEC and KSE, with the latter being relatively neutral and primarily influenced by BSESN and NSEI indices.
\end{itemize}

The presence of substantial spillovers suggests that forecasting volatility time series as a multivariate system could provide benefits. In the context of the mfOU process, this task could be easily achieved using rules for conditional expectations of Gaussian vectors (as, for example, in \cite{BLP21}). This approach is successfully applied in \textcite{bibinger2025modeling} for forecasting multivariate realized volatility with the mfBm. The authors impose the constraint $\eta_{i,j}=0,\ i,j=1,\dots,N$, which rules out asymmetries in the cross-covariance function and is supported empirically in their dataset, consisting in 7 Dow Jones 30 stocks. In our dataset, where asymmetries are present, a similar method should apply, even without imposing the symmetry constraint.

\section{Conclusion}\label{conclusion}

Motivated by recent advances in volatility modeling, we introduce a multivariate model for rough volatility. Our objective is to extend the widely-used rough fractional stochastic volatility model of \textcite{GJR18} to a multivariate framework and empirically study a system of historical volatilities. This extension retains consistency with the findings in the univariate case, namely that volatility is rough and mean reverts slowly, while improving our understanding of volatility dynamics through the analysis of cross-covariances and spillover effects, which are descriptive of how volatility evolves across markets. These results can have implications in risk management and forecasting techniques in finance. 

The results in this paper could be further developed in several directions. From a methodological perspective, it would be interesting to explore how fOU processes capture volatility dynamics at different time scales. From an applied perspective, it would be valuable to study a larger spot-volatility system that exhibits spillover effects in intraday dynamics or, even more importantly, to extend the empirical analysis with rolling-window parameter estimation, which could then be used for dynamic forecasting and spillover analyses.    

In conclusion, by proposing the multivariate fractional Ornstein-Uhlenbeck process for modeling log-volatility time series, we confirm the increasing evidence that rough processes effectively model volatility dynamics and extend our understanding of these dynamics to a multidimensional context, with potential for further applications.

\appendix

\section{Asymptotic theory for the GMM estimator}\label{a-asymptotic-theory-for-the-mde-estimator}

\emph{Proof of Proposition \ref{theo:mde} - (I):} By definition of the GMM estimator, \(\hat\theta_n\) satisfies the first
order condition
\begin{equation*}
\nabla\mathcal{T}_n\left(\theta\right){\huge|} _{\theta=\hat\theta_n}=0,
\end{equation*}
where \(\nabla\mathcal{T}_n\left(\theta\right)\) denotes
the gradient of \(\mathcal{T}_n\left(\theta\right)\).Using a first-order Taylor expansion around \(\theta_0\), the true value
of the parameter, with Lagrange form of the remainder, we have
\begin{equation*}
\nabla\mathcal{T}_n\left(\theta\right) _{\theta=\hat \theta_n}=\nabla\mathcal{T}_n\left(\theta\right){\huge|} _{\theta=\theta_0}+\nabla^2\mathcal{T}_n\left(\theta\right){\huge|} _{\theta= \theta^\star}\left(\hat\theta_n-\theta_0\right)=0,
\end{equation*} 
where \(\theta^\star\) lies between \(\theta_0\) and \(\hat\theta_n\), and \(\nabla^2\mathcal{T}_n\left(\theta\right)\) indicates the Hessian of \(\mathcal{T}\left(\theta\right)\).
Solving for \(\hat\theta_n-\theta_0\), we get 
\begin{equation}
\hat\theta_n-\theta_0=-\left(\nabla^2\mathcal{T}_n\left(\theta\right)\right)^{-1}{\huge|} _{\theta=\theta^\star}\nabla\mathcal{T}_n\left(\theta\right){\huge|} _{\theta=\theta_0}.
\label{eq:est_err}
\end{equation} 
Given the differentiability of \(\gamma_{i,j;n}^k(\theta)\) when \(H_{i,j}\ne 1\), we can explicitly calculate the terms in \eqref{eq:est_err} as 
\begin{equation}
\nabla\mathcal{T}_n\left(\theta\right)=-2J_\gamma^T\mathcal{W}_n\left(\hat\gamma_n-\gamma(\theta)\right),
\label{eq:ddt}
\end{equation} 
and 
\begin{equation}
\nabla^2\mathcal{T}_n\left(\theta\right)=2J_\gamma^T\mathcal{W}_nJ_\gamma+o\left(\hat\gamma_n-\gamma(\theta)\right),
\label{eq:d2dt2}
\end{equation} 
where \(J_\gamma\) is the Jacobian matrix of \(\hat\gamma_n-\gamma(\theta)\) and \(o\left(\hat\gamma_n-\gamma(\theta)\right)\) represents a term containing the second derivatives of \(\gamma(\theta)\), which are bounded in a neighborhood of $\hat{\theta}_n$, times \(\hat\gamma_n-\gamma(\theta)\). 
\textcite{dgp1} show that \(Y_t\) is ergodic and consequently that 
\begin{equation*}
\hat\gamma_{i,j;n}^k \overset{p}\to \gamma_{i,j}(k),\ \ \text{as}\ n\to\infty,
\end{equation*} 
which implies that
\begin{equation*}
\hat\gamma_n\overset{p}{\to} \gamma(\theta) \ \ \ as\ n\to\infty.
\end{equation*} 
From this we can conclude that
\begin{equation}
\nabla\mathcal{T}_n\left(\theta\right)\overset{p}\to 0 
\quad \mbox{ and } \quad
\nabla^2\mathcal{T}_n\left(\theta\right)\overset{p}{\to}2J_\gamma^T\mathcal{W}J_\gamma,
\label{eq:d2lim}
\end{equation} 
where we have assumed $\mathcal{W}_n\to \mathcal{W}$.\\  
These two convergences imply together that \(\theta^\star\overset{p}{\to}\theta_0\) and \(\hat\theta_n\overset{p}{\to}\theta_0\). 
\\
\emph{Proof of Proposition \ref{theo:mde} - (II):}
In order to establish the central limit theorem for \(\hat \theta_n\), we need to understand the limit distribution of the term in (\ref{eq:ddt}).
It is possible to prove that when \(H_i+H_j<\frac{3}{2}\), \(i,j=1,\dots,N,\ \forall k\) 
\begin{equation}
\sqrt{n}\left(\frac{1}{n}\sum_{t=1}^{n-k}Y_{t+k}^iY_t^j-\gamma_{i,j}(k)\right)\overset{d}{\to}N(0,\sigma_{ij}^2(k)),
\label{eq:clt_arc}
\end{equation} 
where 
\[
\sigma_{ij}^2(k)=\text{Var}\left(Y^i_{t+k} Y_t^j\right)+2\sum_{s=0}^\infty\text{Cov}\left(Y^i_{t+k} Y_t^j,Y^i_{t+k+s} Y_{t+k}^j\right).
\] 
The previous result can be extended to the vector \(\hat\gamma_n-\gamma(\theta)\), as 
\begin{equation}
\sqrt{n} \left(\hat\gamma_n-\gamma(\theta)\right)\overset{d}{\to}N(0,\Gamma),
\label{eq:clt_vec}
\end{equation} 
where the diagonal elements of \(\Gamma\) would coincide with \(\sigma_{ij}^2(k)\) above, \(i,j=1,\dots,N,\ k\in\mathcal{L}\), and the additional off-diagonal covariance terms would be in a similar form. The results in (\ref{eq:clt_arc}) and (\ref{eq:clt_vec}) are deduced by Theorem 3.6 in \textcite{dgp1}. See also Theorem 4 in \textcite{arcones1994limit}. Therefore, it is possible to conclude that
\begin{equation*}
\sqrt{n} \nabla\mathcal{T}_n\left(\theta\right) = -\sqrt{n}2J_\gamma^T\mathcal{W}_n\left(\hat\gamma_n-\gamma(\theta)\right)\overset{d}{\to}N(0,4J_\gamma^T\mathcal{W}\Gamma \mathcal{W}J_\gamma)
\end{equation*} 
which together with \eqref{eq:est_err}, \eqref{eq:d2lim}, and Slutsky's Theorem establishes (II).

\section{Dataset description}\label{b-dataset}
\begin{figure}
    \centering 
    \includegraphics[width=0.92\textwidth]{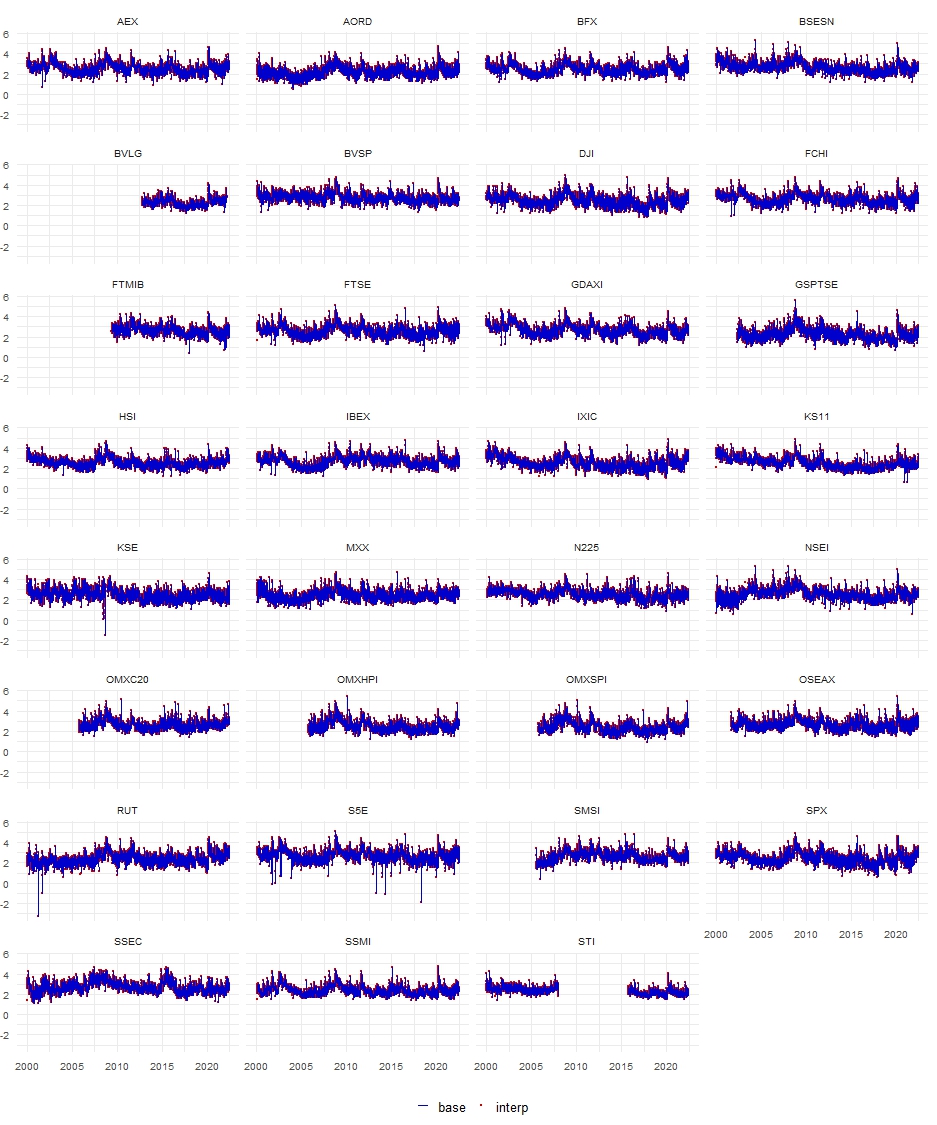} 
    \caption{Time series of $\log\left(100\sqrt{RV\times252}\right)$ for the entire sample, where $RV$ is the realized variance from 5-minute price increments provided in the Oxford-Man realized library. Blue lines represent the original time series, whereas red dots represent the AR(5) interpolation of sparse missing values. Notice how the symbols not included in the empirical analysis (see Table \ref{tab:db}) display prolonged periods of missing values.}
    \label{fig:db} 
\end{figure}

Table \ref{tab:db} and Figure \ref{fig:db} provide a description of the dataset, see the captions for details.

\begin{table}
    \centering
    \caption{Description of the dataset utilized in the empirical analysis obtained from the Oxford-Man realized library. Log-volatilities are computed as $\log\left(100\sqrt{RV\times252}\right)$. The columns Missing and Zeros report the count of days with missing values and zero volatility, respectively. Descriptive statistics include the in-sample average (Mean), standard deviation (SD), minimum (Min), median (Med), and maximum (Max) values. Included specifies wether the time series is included in the multivariate system under analysis or not. }
    \begin{tabular}{llccccccccc}
        \toprule
        \textbf{Symbol} & \textbf{Country} & \multicolumn{2}{c}{\textbf{Count}} & \multicolumn{5}{c}{\textbf{Descriptive Statistics}} & \textbf{Included} \\
        \cmidrule(lr){3-4} \cmidrule(lr){5-9}
        & & \textbf{Missing} & \textbf{Zeros} & \textbf{Mean} & \textbf{SD} & \textbf{Min} & \textbf{Median} & \textbf{Max} & \\
        \midrule
        AEX    & Netherlands & 159  & 0 & 2.53 & 0.51 & 0.69 & 2.48 & 4.63 & Yes \\
        AORD   & Australia   & 218  & 0 & 2.14 & 0.48 & 0.51 & 2.09 & 4.70 & Yes \\
        BFX    & Belgium     & 161  & 0 & 2.46 & 0.47 & 1.16 & 2.42 & 4.57 & Yes \\
        BSESN  & India       & 314  & 9 & 2.64 & 0.49 & 1.12 & 2.59 & 5.27 & Yes \\
        BVLG   & Peru        & 3495 & 1 & 2.27 & 0.38 & 1.17 & 2.26 & 4.14 & No \\
        BVSP   & Brazil      & 363  & 0 & 2.76 & 0.41 & 1.28 & 2.73 & 4.80 & Yes \\
        DJI    & USA         & 264  & 0 & 2.44 & 0.56 & 0.79 & 2.41 & 4.99 & Yes \\
        FCHI   & France      & 157  & 0 & 2.63 & 0.50 & 0.97 & 2.61 & 4.73 & Yes \\
        FTMIB  & Italy       & 2579 & 0 & 2.60 & 0.45 & 0.33 & 2.59 & 4.42 & No \\
        FTSE   & UK          & 226  & 0 & 2.54 & 0.50 & 0.61 & 2.48 & 5.10 & Yes \\
        GDAXI  & Germany     & 196  & 0 & 2.68 & 0.52 & 1.17 & 2.64 & 4.80 & Yes \\
        GSPTSE & Canada      & 857  & 0 & 2.21 & 0.55 & 0.67 & 2.14 & 5.61 & No \\
        HSI    & Hong Kong   & 389  & 0 & 2.54 & 0.42 & 1.20 & 2.49 & 4.65 & Yes \\
        IBEX   & Spain       & 194  & 0 & 2.70 & 0.47 & 1.17 & 2.71 & 4.77 & Yes \\
        IXIC   & USA         & 257  & 0 & 2.54 & 0.55 & 0.95 & 2.48 & 4.81 & Yes \\
        KS11   & South Korea & 362  & 0 & 2.54 & 0.51 & 0.61 & 2.48 & 4.81 & Yes \\
        KSE    & Pakistan    & 416  & 0 & 2.47 & 0.52 & -1.52 & 2.43 & 4.57 & Yes \\
        MXX    & Mexico      & 257  & 0 & 2.40 & 0.45 & 1.10 & 2.35 & 4.74 & Yes \\
        N225   & Japan       & 434  & 0 & 2.52 & 0.47 & 0.83 & 2.50 & 4.59 & Yes \\
        NSEI   & India       & 318  & 12 & 2.53 & 0.52 & 0.53 & 2.49 & 5.32 & Yes \\
        OMXC20 & Denmark     & 1740 & 0 & 2.59 & 0.44 & 1.43 & 2.52 & 5.17 & No \\
        OMXHPI & Finland     & 1699 & 0 & 2.45 & 0.50 & 1.16 & 2.37 & 5.47 & No \\
        OMXSPI & Sweden      & 1699 & 0 & 2.41 & 0.51 & 0.90 & 2.33 & 5.09 & No \\
        OSEAX  & Norway      & 707  & 0 & 2.59 & 0.48 & 1.41 & 2.53 & 5.41 & No \\
        RUT    & USA         & 260  & 0 & 2.35 & 0.51 & -3.32 & 2.30 & 4.53 & Yes \\
        SMSI   & Switzerland & 1565 & 0 & 2.62 & 0.48 & 0.38 & 2.61 & 4.82 & No \\
        SPX    & USA         & 259  & 0 & 2.43 & 0.57 & 0.56 & 2.39 & 4.94 & Yes \\
        SSEC   & China       & 467  & 0 & 2.67 & 0.53 & 1.09 & 2.61 & 4.64 & Yes \\
        SSMI   & Switzerland & 259  & 0 & 2.40 & 0.45 & 1.36 & 2.31 & 4.77 & Yes \\
        STI    & Singapore   & 2203 & 1 & 2.33 & 0.34 & 1.39 & 2.30 & 4.22 & No \\
        S5E & Europe    & 173  & 1 & 2.68 & 0.53 & -1.88 & 2.66 & 5.11 & Yes \\
        \bottomrule
    \end{tabular}
    \label{tab:db}
\end{table}

\section{Empirics of marginal components}\label{d-empirics-of-marginal-components}

\begin{figure}
    \centering 
    \includegraphics[width=\textwidth]{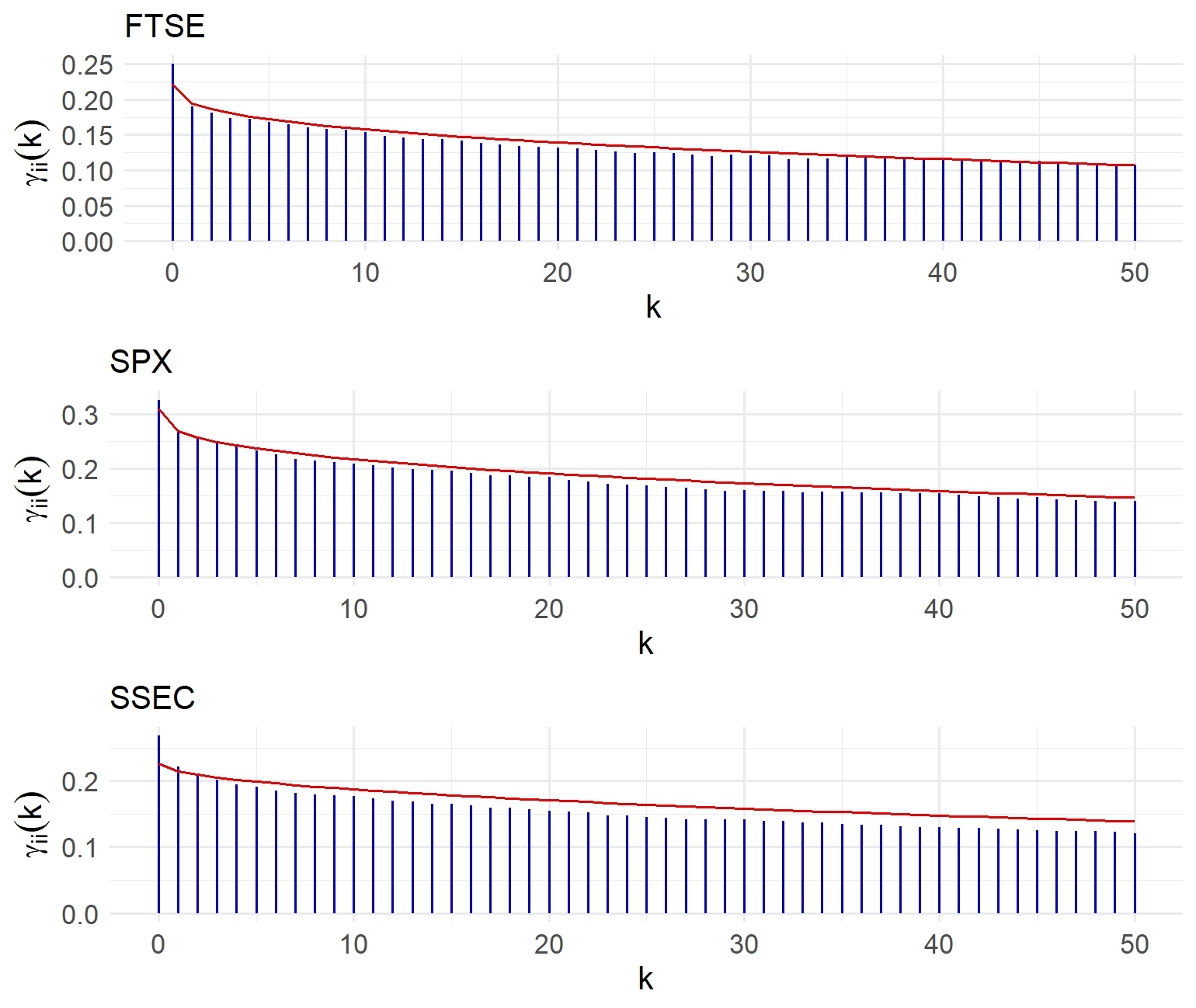}
    \caption{
Empirical autocovariances of log-realized volatilities as blue bars, alongside the theoretical autocovariances from our model, indicated by the red curves and based on the estimated parameters from Table \ref{tab:univ}. The panels, arranged from top to bottom, correspond to the indices FTSE, SPX, and SSEC.}
    \label{fig:acfe} 
\end{figure}

\begin{figure}
    \centering 
    \includegraphics[width=\textwidth]{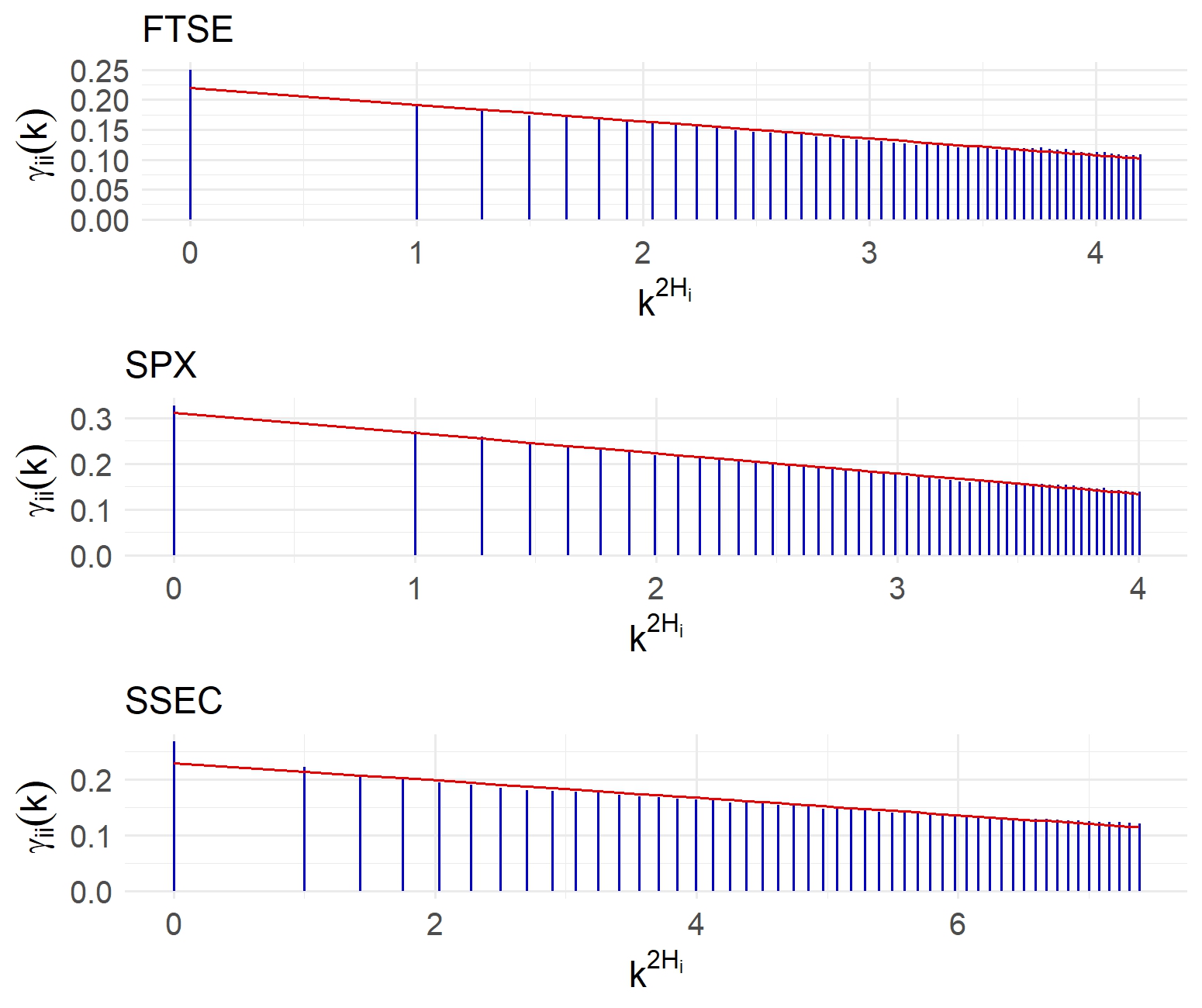}
    \caption{Empirical autocovariances of log-realized volatilities plotted against a suitable power of the lag (given by twice the Hurst exponent) as blue bars, alongside the best linear fits, indicated by the red lines. The panels, arranged from top to bottom, correspond to the indices FTSE, SPX, and SSEC.}
    \label{fig:acfl} 
\end{figure}

\begin{figure}
    \centering 
    \includegraphics[width=\textwidth]{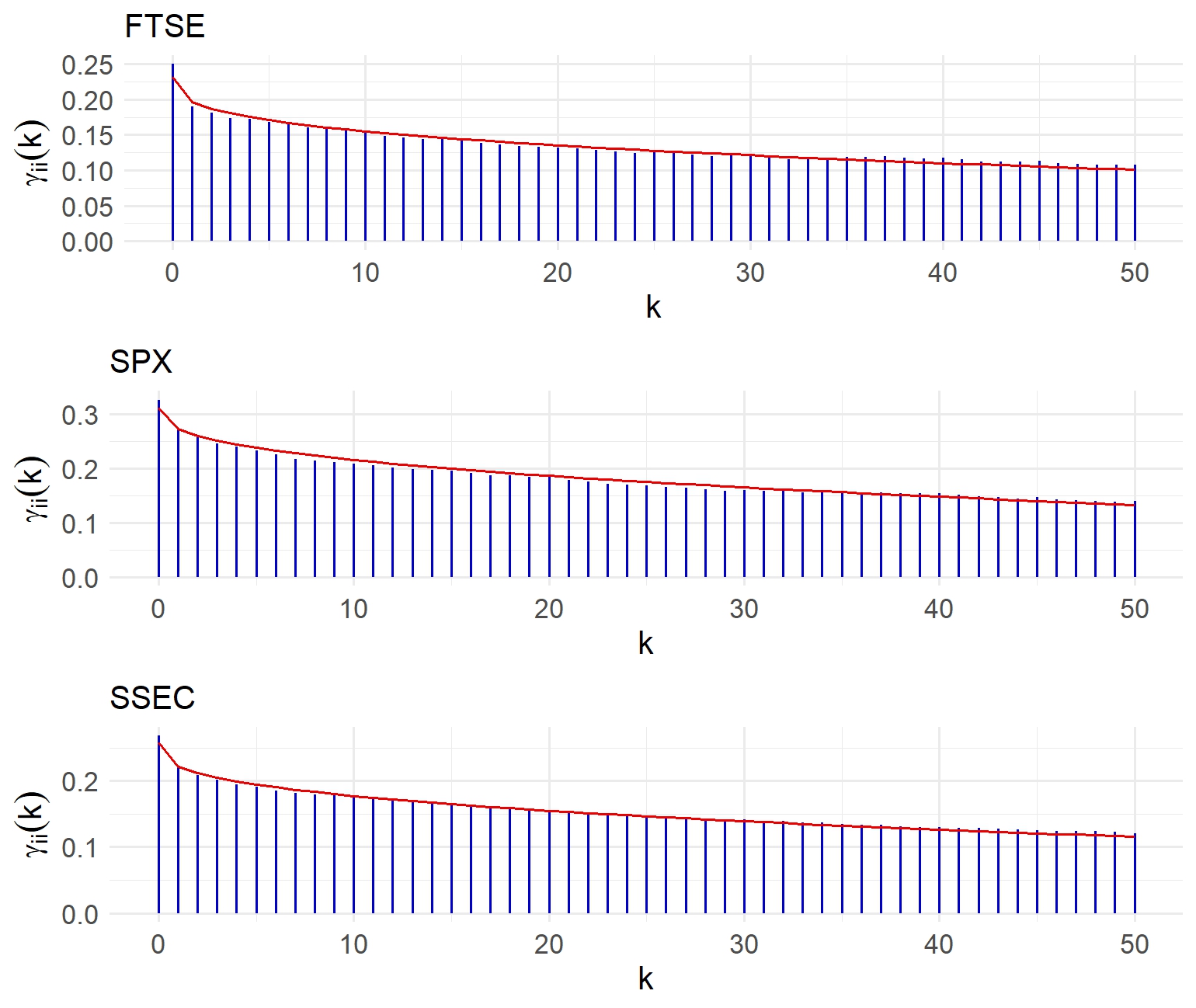}
    \caption{Empirical autocovariances of log-realized volatilities as blue bars, alongside the theoretical autocovariances from our model when $\alpha_i\to 0$, indicated by the red curves and based on the estimated parameters (available in the \href{https://ranieridugo.github.io/mfou_vol_appendix}{Online Appendix}). The panels, arranged from top to bottom, correspond to the indices FTSE, SPX, and SSEC.}
    \label{fig:acfa} 
\end{figure}

\begin{figure}
    \centering 
    \includegraphics[width=\textwidth]{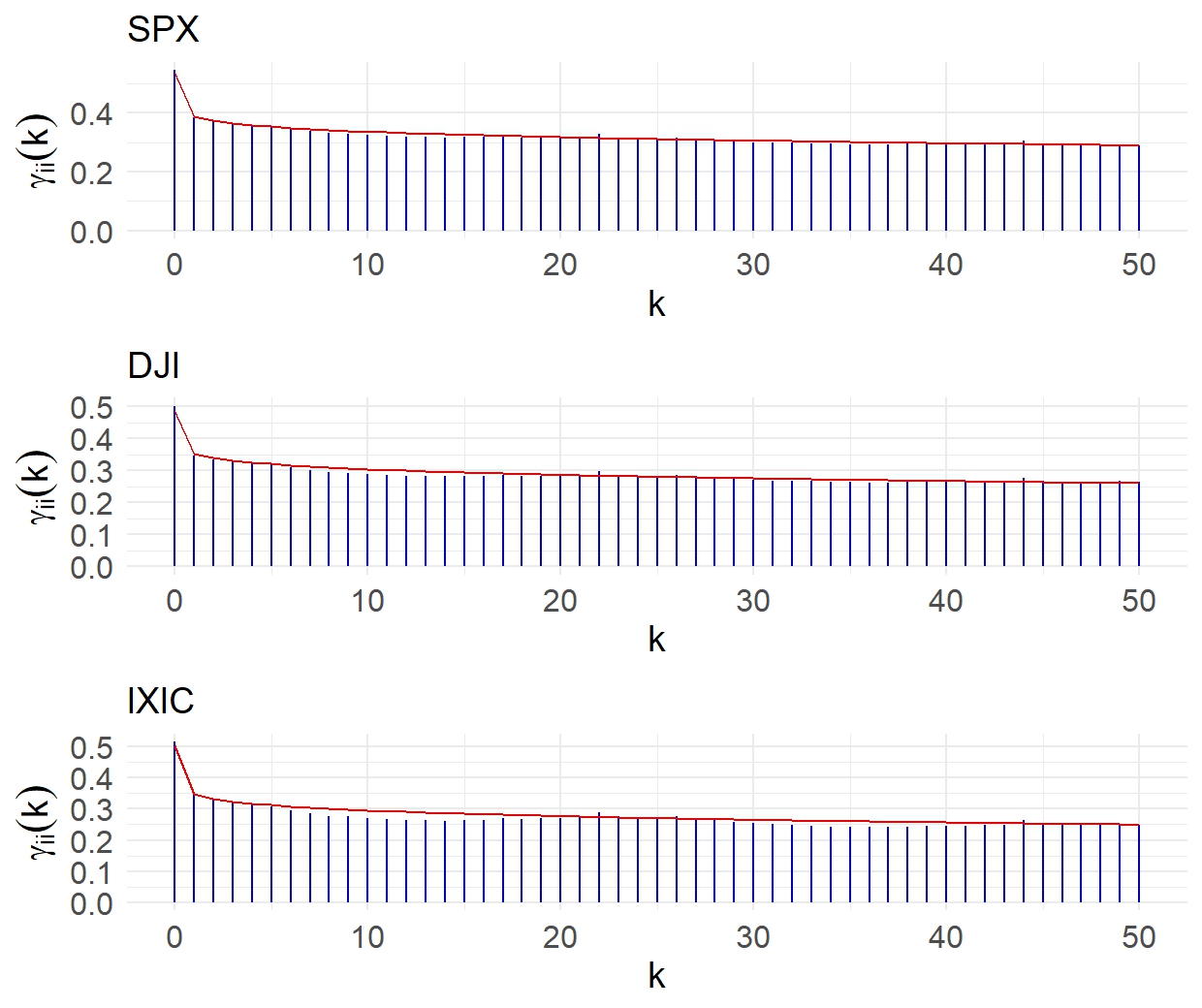}
    \caption{Empirical autocovariances of log-spot volatilities as blue bars, alongside the theoretical autocovariances from our model, indicated by the red curves and based on the estimated parameters from Table \ref{tab:spot-est}. The panels, arranged from top to bottom, correspond to the indices SPX, DJI, and IXIC}
    \label{fig:acfe_spot} 
\end{figure}

The Figures \ref{fig:acfe}, \ref{fig:acfl}, \ref{fig:acfa}, and \ref{fig:acfe_spot} provide additional results to the analysis presented in Section \ref{empirical-analysis}, focusing on the univariate marginal components. They display the same results shown in Figures \ref{fig:ccfe}, \ref{fig:ccfl}, \ref{fig:ccfa}, and \ref{fig:ccfe_spot} but in terms of the autocovariances of the marginal components. They corroborate the evidence that our model provides a good fit to log-realized volatility time series.

\section{Spillovers methodology}\label{e-spillovers-methodology}

Starting from a discrete-time moving average representation of a discrete time process \(Y_t\in \mathbb{R}^N\),
\begin{equation}
Y_t=\sum_{k=0}^\infty A_i\varepsilon_{t-k},
\label{eq:mad}
\end{equation} 
where \(\varepsilon_t\overset{iid}{\sim}N(0,\Sigma)\) and \(A_i\in\mathbb{R}^{N\times N}\) such that the representation (\ref{eq:mad}) is well-defined, \textcite{pesaran1997working} and \textcite{pesaran1998generalized} show that 
\begin{equation}
\begin{aligned}
\psi_{i,j}(h)=&
\frac{\mathbb{E}\left[\left(Y_{t+h}^i-\mathbb{E}\left[Y_{t+h}^i|\mathcal{F}_{t-1}\right]\right)^2\right]-
\mathbb{E}\left[\left(Y_{t+h}^i-\mathbb{E}\left[Y_{t+h}^i|\mathcal{F}_{t-1},\left(\varepsilon^j_{t+i}\right)_{i=0}^{h}\right]\right)^2\right]}
{\mathbb{E}\left[\left(Y_{t+h}^i-\mathbb{E}\left[Y_{t+h}^i|\mathcal{F}_{t-1}\right]\right)^2\right]} \\
=&\frac{\sqrt{\Sigma_{i, i}}^{-1}\sum_{\ell=0}^h\left(e_i^TA_\ell\Sigma e_j\right)^2}{\sum_{\ell=0}^ne_i^TA_\ell\Sigma A_\ell^Te_i},
\label{eq:fevdd}
\end{aligned}
\end{equation} 
where \(\mathcal{F}_{t-1}\) contains information regarding all the past innovations \(\varepsilon_{t-i},\ i = 1,2,\dots\). The definition for \(\psi_{i,j}(h)\) is used in \textcite{diebold2012better} to construct spillover indices and analyse them. Therefore, we frame the mfOU process in the same setting as above in order to conduct a similar analysis. 
Consider the representation of the causal mfOU (\cite{dgp1}): 
\begin{equation}
Y_t=\int_{-\infty}^tK(t,s)MdW_s,
\label{eq:mac}
\end{equation} 
where the kernel \(K(t,s):\mathbb{R}^2\to\mathbb{R}^N\) is diagonal with components 
\begin{equation*}
K_i(t,s)=\nu_i\left((t-s)_+^{H_i-\frac{1}{2}}-(-s)_+^{H_i-\frac{1}{2}}-\alpha_i\int_s^te^{-\alpha_i(t-u)}\left((u-s)_+^{H_i-\frac{1}{2}}-(-s)_+^{H_i-\frac{1}{2}}\right)du\right),
\end{equation*} 
\(i=1,\dots,N\), \(M\) is an \(N\times N\) matrix such that \(MM^T=P\) (from \cite{ACLP10}), with
\[
P_{i,j}=\frac{\sin\left(\pi(H_i+H_j)\right)}{B\left(H_i+\frac{1}{2},H_j+\frac{1}{2}\right)\left(\cos\left(\pi H_i\right)+\cos\left(\pi H_j\right)\right)}\rho_{i,j},
\] 
where \(B(x,y),\ x,y>0\), denotes the Beta function, and \(W_t\) is a standard N-dimensional Brownian motion. Note that
the representation in (\ref{eq:mac}) introduces a modification compared to that in \textcite{dgp1}, as it factors \(M\) out of the kernel. This factorization enables the separation of the dependence on time from that between components, aligning the model with the framework in (\ref{eq:mad}).
Equivalently, 
\begin{equation*}
\begin{aligned}
Y_t^i=&\int_{-\infty}^t K_i(t,s)\sum_{j=1}^NM_{i,j}dW_s^j \\
=&\sqrt{\frac{\sin(\pi H_i)}{B\left(H_i+\frac{1}{2},H_i+\frac{1}{2}\right)}}\int_{-\infty}^t K_i(t,s)\sum_{j=1}^N\sqrt{\frac{B\left(H_i+\frac{1}{2},H_i+\frac{1}{2}\right)}{\sin(\pi H_i)}}M_{i,j}dW_s^j \\
=&\sqrt{\frac{\sin(\pi H_i)}{B\left(H_i+\frac{1}{2},H_i+\frac{1}{2}\right)}}\int_{-\infty}^t K_i(t,s)\sum_{j=1}^Nd\overline{W}_s^j
\end{aligned} 
\end{equation*} 
where \(\overline{W}\) is a Brownian motion with covariance matrix \( G\), where 
\[
 G_{i,j}=\sqrt{\frac{B(H_i+\frac{1}{2},H_i+\frac{1}{2})B(H_j+\frac{1}{2},H_j+\frac{1}{2})}{\sin\left(\pi H_i\right)\sin\left(\pi H_j\right)}}\frac{1}{B(H_i+\frac{1}{2},H_j+\frac{1}{2})}\frac{\sin\left(\pi(H_i+H_j)\right)}{\left(\cos\left(\pi H_i\right)+\cos\left(\pi H_j\right)\right)}\rho_{i,j}.
\] 
Note that $\overline{W}_\cdot^j$ is a Brownian motion with unit variance, for any $j$.
We can rewrite 
\[
Y_t=\int_0^t\overline K(t-s)d\overline{W}_s+\int_{-\infty}^0K^*(t,s)d\overline{W}_s,
\]
where \(K^*(t,s)=K(t,s)\sqrt{\sin(\pi H_i)/ B (H_i+1/2,H_i+1/2)}\) and
\[
\overline K(t-s)_{i,j}=\nu_i\sqrt{\frac{\sin(\pi H_i)}{B\left(H_i+\frac{1}{2},H_i+\frac{1}{2}\right)}}\left((t-s)^{H_i-\frac{1}{2}}-\alpha_i\int_s^te^{-\alpha_i(t-u)}\left((u-s)^{H_i-\frac{1}{2}}\right)du\right).\]
One can check that this dependens only on \(t-s\) with simple change of variables.  
At this point we can approximate the process using a left point discretization scheme over a grid with uniform mesh \(\Delta\), introducing \(\varepsilon_{l}:=\Delta_l\overline{W}=\overline{W}_{l\Delta}-\overline{W}_{(l-1)\Delta}\), as 
\begin{equation}
\begin{aligned}
Y_t\approx&\sum_{l=1}^{\frac{t}{\Delta}}\overline K\left(t-(l-1)\Delta \right) \Delta_l\overline{W}+\sum_{l=-\infty}^{0} K^*\left(t,(l-1)\Delta \right)\Delta_l\overline{W} \\
=&\sum_{k=0}^{\frac{t}{\Delta}-1}\overline K\left((k+1)\Delta\right)\varepsilon_{\frac{t}{\Delta}-k} + \sum_{k=\frac{t}{\Delta}}^{+\infty}K^*(t,t-(k+1)\Delta)\varepsilon_{\frac{t}{\Delta}-k}\\
=&\sum_{k=0}^{t-1}A_k\varepsilon_{t-k}+\sum_{k=t}^{\infty}B_{k,t}\varepsilon_{t-k},
\label{eq:mfoud}
\end{aligned}
\end{equation} 
where we fix \(t:=\frac{t}{\Delta},\ k+l=t, \ A_k=\overline K\left((k+1)\Delta\right), \ \text{and}\ \ B_{k,t}=K^*\left(t,t-(k+1)\Delta\right)\).
When \(t\ge h>0\), the calculation in (\ref{eq:fevdd}) delivers the same result for the discretized mfOU process in (\ref{eq:mfoud}) as for the process in (\ref{eq:mad}), due to the common convolution term \(\sum_{k=0}^{t-1}A_k\varepsilon_{t-k}\), and therefore
\begin{equation}
\begin{aligned}
\psi_{i,j}(h)=&\frac{\Delta \sqrt{\Sigma_{j,j}}^{-1}\sum_{l=0}^{h-1}\left(e_i^T\overline K\left((l+1)\Delta\right) G e_j\right)^2}{\sum_{l=0}^{h-1}\left(e_i^T\overline K\left((l+1)\Delta\right) G \overline K\left((l+1)\Delta\right)^T e_i\right)} \\
=& \frac{\Delta\sum_{l=0}^{h-1}\overline K_{i,j}^2\left((l+1)\Delta\right) G^2_{i,j}}{\sqrt{\Sigma_{j,j}}\sum_{l=0}^{h-1}\overline K_{i,j}^2\left((l+1)\Delta\right) G_{i,i}} \\
=&\frac{\Delta G_{i,j}^2}{\sqrt{ G_{j,j}} G_{i,i}},
\end{aligned}
\end{equation} and \begin{equation}
\widetilde\psi_{i,j}=\frac{\psi_{i,j}}{\sum_{j=1}^N\psi_{i,j}}\\
=\frac{ G_{i,j}^2/\sqrt{ G_{j,j}}}{\sum_{m=1}^N G_{i,m}^2/\sqrt{ G_{m,m}}}.
\label{eq:psiij}
\end{equation}
\begin{remark}
The result in (\ref{eq:psiij}) is obtained under the assumption that $\varepsilon_{t+i}^j,\ i=0,\dots,n,\ j=1,\dots,N$ follows a white noise process in the moving average representation (\ref{eq:mad}). A similar spillover analysis could be performed by relaxing the i.i.d. assumption and, instead of using the moving average representation in terms of white noises, simply conditioning on $\varepsilon_{t+i}^j,\ i=1,\dots,n,\ j=1,\dots,N$ being fractional Gaussian noise innovations. Preliminary results suggest qualitatively similar outcomes, though with a dependency on the forecast horizon and a larger computational burden. 
\end{remark}

\section{Additional material}\label{additional-material}

The Online Appendix, available at \href{https://ranieridugo.github.io/mfou_vol_appendix}{ranieridugo.github.io/mfou\_vol\_appendix}, 
provides additional empirical results in terms of parameter estimates and covariance fit for the components of the system not reported here, as well as the small-alpha (slow mean reversion) and causal regimes. Working code to simulate, estimate, forecast, and analyze volatility spillovers using the mfOU process is available at \href{https://github.com/ranieridugo/mfou}{github.com/ranieridugo/mfou}.

\paragraph*{Disclosure of interest.}
We declare no conflict of interest.

\paragraph*{Data Availability Statement.}
The realized-volatility analysis in this paper uses a dataset downloaded in June 2022 from \href{https://oxford-man.ox.ac.uk/research/realized-library}{https://oxford-man.ox.ac.uk/research/realized-library} in June 2022. A similar dataset is openly available at \href{https://dachxiu.chicagobooth.edu}{https://dachxiu.chicagobooth.edu}. The spot-volatility analysis relies on a dataset purchased from \href{https://firstratedata.com/b/8/us-index-historic-intraday}{https://firstratedata.com/b/8/us-index-historic-intraday}. The code on which the analysis is based is openly available at \href{https://github.com/ranieridugo/mfou}{https://github.com/ranieridugo/mfou}.

\printbibliography

\end{document}